\documentclass[preprints,article,accept,pdftex,moreauthors]{Definitions/mdpi} 
\firstpage{1} 
\pubvolume{1}
\issuenum{1}
\articlenumber{0}
\pubyear{2026}
\copyrightyear{2026}
\datereceived{ } 
\daterevised{ } 
\dateaccepted{ } 
\datepublished{ } 
\hreflink{https://doi.org/} 

\Title{Determination of Physical Height Differences from Time Transfer via the ACES Mission – A Simulation Study}

\TitleCitation{Determination of Physical Height Differences from Time Transfer via the ACES Mission – A Simulation Study}

\Author{Klarissa Emma Lachmann $^{1,*}$\orcidA{}, Jürgen Müller $^{1}$\orcidB{}, Peter Vollmair $^{2}$\orcidC{} and Anja Schlicht $^{2}$}

\AuthorNames{Klarissa Emma Lachmann, Jürgen Müller, Peter Vollmair and Anja Schlicht}

\isAPAStyle{%
       \AuthorCitation{Lachmann, K. E., Müller, J., Vollmair, P. \& Schlicht, A.}
         }{%
        \isChicagoStyle{%
        \AuthorCitation{Lastname, Firstname, Firstname Lastname, and Firstname Lastname.}
        }{
        \AuthorCitation{Lachmann, K.E.; Müller, J.; Vollmair, P.; Schlicht, A.}
        }
}

\address{%
$^{1}$ \quad Institut für Erdmessung, Leibniz Universität Hannover, 30167 Hanover, Germany; mueller@ife.uni-hannover.de 
\\
$^{2}$ \quad Forschungseinrichtung Satellitengeodaesie, Technical University of Munich, 80333 Munich, Germany; peter.vollmair@tum.de (P.V.); anja.schlicht@tum.de (A.S.)}

\corres{Correspondence: lachmann@ife.uni-hannover.de}

\addhighlights{
What are the main findings?
\begin{itemize}[leftmargin=*]
    \item Satellite-based optical time transfer enables the determination of physical height differences between two distant locations at the decimeter level within just a few days of observation.
    \item Including non-common view observations of the satellite from the ground stations increases the number of measurements while maintaining an accuracy comparable to that of common view configurations.
\end{itemize}

\noindent What is the implication of the main finding?
\begin{itemize}[leftmargin=*]
    \item Satellite-based time transfer provides an approach for determining physical height differences over large distances without requiring dedicated terrestrial optical fiber links between the clock sites.
    \item The developed analysis strategy provides a foundation for utilizing future ACES-like time-transfer observations in relativistic geodesy.
\end{itemize}
}

\abstract{
The determination of physical height differences using highly stable atomic clocks has emerged as a novel approach in relativistic geodesy, exploiting the gravitational redshift as a direct observable of geopotential differences. In this study, we investigate the feasibility of satellite-based clock comparisons using the Atomic Clock Ensemble in Space (ACES) onboard the International Space Station, which enables time transfer via microwave (MWL) and optical (ELT) links. Since operational optical data are not yet available, a comprehensive full-scale simulation of realistic ACES observation scenarios is performed, including detailed noise models of clocks and links. A slope-based estimation method is applied to time series of clock comparisons in order to extract the relativistic redshift signal and derive height differences between the ground stations. The performance of the approach is evaluated for quasi-common view, non-common view, and split non-common view configurations, where the latter divides the observation period into shorter intervals. The results show that optical links enable faster convergence and can achieve height accuracies at the decimeter level within a few days and at the centimeter level over longer periods, while microwave links are more strongly affected by noise and bias contributions. Non-common view processing significantly increases observation availability with only minor loss in accuracy, and the split approach provides robust solutions for larger networks. These findings demonstrate the strong potential of satellite-based clock comparisons as a remote-sensing technique for determining physical height differences on a continental scale.
}


\begin{document}
\section{Motivation}
According to general relativity, the rate of an ideal clock depends on the gravitational potential at its location. Two clocks positioned at distinct heights in the Earth's gravity field therefore tick at different rates. This relative rate difference, known as the gravitational redshift, leads to an accumulated time offset between the clocks that is directly proportional to the geopotential difference between their locations \cite{bib_mul_18, bib_denker_18}. By comparing highly stable clocks at distant sites, this effect can be exploited to determine gravity-related height differences without relying on classical leveling or gravity observations.
With the rapid progress in optical atomic clocks, fractional frequency uncertainties at or below the $10^{-18}$ level have been achieved, corresponding to a potential height resolution on the order of a few centimeters on Earth \cite{bib_bel_21}. Such performance transforms clock comparisons into a powerful new tool for geodesy and Earth observation, enabling the direct measurement of geopotential differences over large distances \cite{bib_vincent_24, bib_mcg_18, bib_fal_14, bib_lis_16, bib_bon_12}.

Traditionally, high-precision comparisons of distant atomic clocks are performed via phase-coherent optical fiber links, which enable direct frequency comparisons with extremely high stability and accuracy. Such links have demonstrated fractional uncertainties at or below $10^{-18}$ over continental distances and have already been used to measure geopotential differences between remote sites \cite{bib_vin_24, bib_lis_16, bib_dro_13, bib_pre_12, bib_rau_14}. However, their applicability is restricted by the availability of dedicated fiber infrastructure and is therefore largely limited to more regional networks.
Satellite-based time and frequency transfer provides a complementary approach that enables comparisons between distant clocks on a global scale without the need for ground-based connections \cite{bib_lev_08}.
In this context, the Atomic Clock Ensemble in Space (ACES) mission onboard the International Space Station (ISS) represents a major milestone for space-based precision timing and fundamental physics experiments \cite{bib_cac_09, bib_cac_07}. ACES combines a cold-atom cesium clock (PHARAO), providing long-term stability, with a space hydrogen maser (SHM) for short-term stability. The mission supports time transfer via the Microwave Link (MWL), a two-way time-transfer system operating in multiple frequency bands, and the European Laser Timing (ELT) experiment, which enables optical time transfer via laser ranging techniques \cite{bib_lau_15}. In contrast to phase-coherent optical fiber links, which allow direct frequency comparisons, the ELT measurements are based on pulsed time-of-flight observations and therefore do not provide a continuous phase observable \cite{bib_sch_09}. This limits the applicability of classical frequency-domain approaches and favors a time-domain treatment. Together, these instruments enable tests of fundamental physics as well as applications in relativistic geodesy. In particular, satellite-mediated clock comparisons allow the determination of geopotential differences between remote locations \cite{bib_cac_09, bib_sch_18, bib_meh_18}.

However, extracting the gravitational redshift signal from satellite time-transfer data remains challenging, as the observations are affected by measurement noise and modeling errors, e.g. those related to the satellite orbit, and are further limited by the restricted visibility of the satellite. Classical common view techniques require simultaneous observations of the satellite at both ground stations, which may not always be feasible. Non-common view methods relax this constraint but demand on satellite clock errors and propagation effects. Robust analysis strategies are therefore required to exploit ACES data for geodetic applications.

Several studies have investigated clock-based geopotential determination using optical clocks and satellite links, demonstrating the theoretical feasibility of centimeter-level height measurements \cite{bib_bon_12, bib_del_18, bib_vincent_24, bib_mul_20}. In practice, however, the performance strongly depends on the stability and noise characteristics of the clocks and links, as well as on the observation geometry and data processing approach. In this work, we investigate a slope-based method for extracting relativistic redshift information from satellite time-transfer observations. Instead of directly analyzing phase or time offsets, the method estimates linear trends in clock desynchronization time series to derive height differences between the ground stations. The approach is applied to simulated ACES observations including both microwave and optical links and realistic noise models for multiple European ground stations.
Since operational data are not yet available, we introduce a comprehensive full-scale simulation of the ACES observation environment. Different observation configurations are considered, including quasi-common view, non-common view, and split non-common view scenarios. Here, the term split non-common view describes  a processing strategy where the non-common view observation period is divided into shorter intervals. This allows us to assess the achievable accuracy under realistic conditions and to evaluate the robustness of the method with respect to incomplete satellite visibility.

By providing a detailed simulation-based assessment, this work serves as a preparatory study for the analysis of forthcoming ACES measurements and demonstrates the potential of satellite-based clock comparisons as a remote-sensing technique for relativistic geodesy. The remainder of this paper is structured as follows. Section~\ref{Sim_Back} introduces the simulation framework, including the observation models and noise characteristics. Section~\ref{sec:Meth} describes the methodology for deriving gravity-related height differences from clock desynchronization, with particular focus on the slope-based estimation approach and the considered observation scenarios. The results obtained for quasi-common view, non-common view, and split non-common view configurations are presented and discussed in Section~\ref{sec:results}. Finally, Section~\ref{sec:conclusion} summarizes the main findings and outlines perspectives for future work.

\section{Simulation Background}
\label{Sim_Back}
\begin{figure}[H]
	\centering
	\includegraphics[width=9.0 cm]{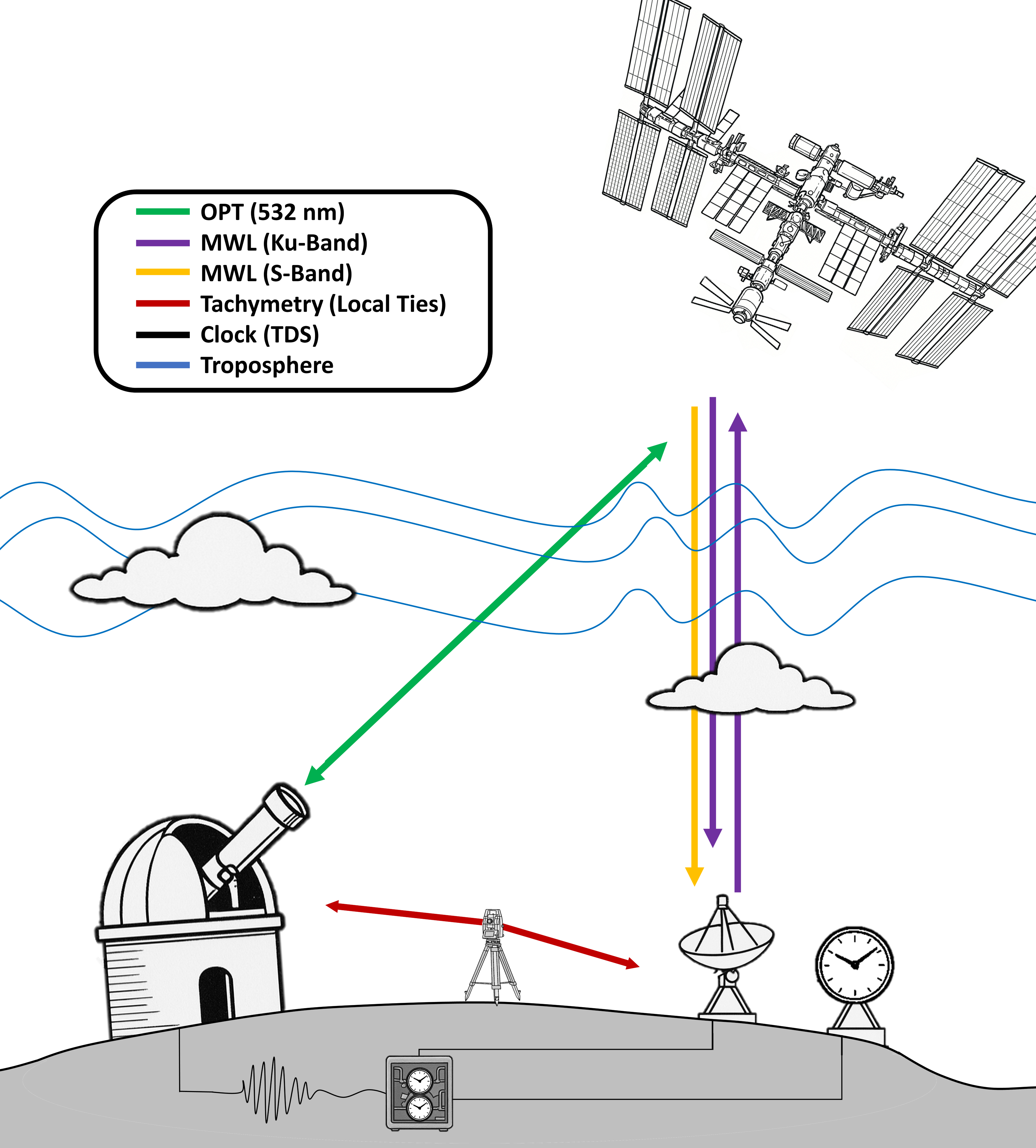}

	\caption{Simulation setup at GOW. Figure from \cite{bib_vol_25}\label{fig_gow_sketch}}
\end{figure}

The data set used in this study is based on the ACES Full-Scale Simulation and Analysis Tool developed by Vollmair et~al (2023) \cite{bib_vol_23} in preparation for the launch of the ACES mission. The requirement for this software was to represent the mission's observation environment as accurately as possible. Figure~\ref{fig_gow_sketch} provides an overview of the typical simulation setup, which is oriented toward the Geodetic Observatory Wettzell (GOW). 
\unskip
The observation environment includes a laser-optical link (OPT) and a microwave link (MWL) for observations between the International Space Station (ISS) and a ground station, in this example, the GOW. At the ground station, both observation techniques are connected to a common clock using a time distribution system (TDS). At the ISS, the optical detector, the microwave transmission, and the receiving system are connected to the ACES clock. A local tie measurement is assumed for the combination of the collocated observation techniques at the ground station. The OPT observations, with a wavelength of $532$ nm, consist of one-way measurements (\ref{eq_theta_1}), in which an optical detector on the ISS registers photons from the SLR station, as well as two-way measurements (\ref{eq_theta_2}) between a laser reflector array (LRA) on the ISS and an SLR station. Laser-based one-way observations are pseudorange observations that have already been assigned to the corresponding laser-based two-way observations. The measurement frequencies are $100$ Hz for the one-way measurements and $300$ Hz for the two-way measurements.
\begin{adjustwidth}{-\extralength}{0cm}
	\begin{linenomath}
		\begin{equation}
			\theta_{1}= \left \lVert \boldsymbol{r}_{\theta_{1}}^{s}(\tau^{s})-\boldsymbol{r}_{\theta} \right 	\rVert-c\cdot(\delta \tau^{s} - \delta \tau_{r})+\gamma_{sha}+\gamma_{sag}+t_{i_{\theta}}+t_{a_{\theta}}+\sigma_{t_{\theta_{1}}}+\sigma_{\theta_{1}}\label{eq_theta_1}
		\end{equation}
	\end{linenomath}
	\begin{linenomath}
		\begin{equation}
			\theta_{2} = 2 \cdot (\left \lVert 	\boldsymbol{r}_{\theta_{2}}^{s}(\tau_{r})-\boldsymbol{r}_{\theta} \right \rVert+\rho_{lra}+\gamma_{sha}+\gamma_{sag}+t_{i_{\theta}}+t_{a_{\theta}}+\sigma_{t_{\theta_{2}}}+\sigma_{\theta_{2}})\label{eq_theta_2}
		\end{equation}
	\end{linenomath}
\end{adjustwidth}
The first term describes the range part of the observation equation, where $\boldsymbol{r}^{s}_{\theta_{1}}$ represents the orbit of the optical detector, $\boldsymbol{r}^{s}_{\theta_{2}}$ represents the orbit of the SLR reflector, and $\boldsymbol{r}_{\theta}$ represents the position of the telescope on ground. The clock term is described as the difference between the satellite- $\delta \tau^{s}$ and the ground clock error $\delta \tau_{r}$. The relativistic influences include the Shapiro $\gamma_{sha}$ and the Sagnac effect $\gamma_{sag}$. The influence of the troposphere on the signal propagation time is described by an azimuth-independent term $t_{i_{\theta}}$, an azimuth-dependent term $t_{a_{\theta}}$, and tropospheric fluctuations $\sigma_{t_{\theta}}$. In addition, $\sigma_{\theta}$ describes the instrument noise and the delay term of the laser reflector array (LRA) $\rho_{lra}$ is added for SLR observations.\\
The MWL observations consist of code and phase observations from two downlinks (\ref{eq_phi_1}) in the S- and Ku bands, as well as from an uplink (\ref{eq_phi_2}) also in the Ku band. The measurement frequency is $12.5$ Hz. With the two downlink observations, the ionospheric influence on the signal propagation time of the microwave observations can be minimized. The observation equations of the two downlinks do not differ and are therefore shown as one. Also, only the observation equations of the phase observations are shown, since the code observations differ only in the sign of the ionospheric signal delay and in the ambiguities.
\begin{adjustwidth}{-\extralength}{0cm}
\begin{linenomath}
	\begin{equation}
		\phi_{1}= \left \lVert \boldsymbol{r}_{\phi_{1}}^{s}(\tau_{r})-\boldsymbol{r}_{\phi} \right \rVert+c\cdot(\delta \tau^{s} - \delta \tau_{r})+\gamma_{sha}+\gamma_{sag_{1}}-i_{1}+b_{1}+N_{1} \cdot \lambda_{1}+t_{i_{\phi_{1}}}+t_{a_{\phi_{1}}}+\sigma_{t_{\phi_{1}}}+\sigma_{\phi_{1}}\label{eq_phi_1}
	\end{equation}
\end{linenomath}
\begin{linenomath}
	\begin{equation}
		\phi_{2}= \left \lVert \boldsymbol{r}_{\phi_{2}}^{s}(\tau^{s})-\boldsymbol{r}_{\phi} \right \rVert-c\cdot(\delta \tau^{s} - \delta \tau_{r})+\gamma_{sha}+ \gamma_{sag_{2}}- i_{2}+ b_{2}+N_{2} \cdot \lambda_{2}+ t_{i_{\phi_{2}}}+ t_{a_{\phi_{2}}}+\sigma_{t_{\phi_{2}}}+\sigma_{\phi_{2}}\label{eq_phi_2}
	\end{equation}
\end{linenomath}
\end{adjustwidth}
In addition to the parameters already mentioned, the MWL observation equations also include terms for the frequency-dependent ionosphere-induced propagation delay $i_{1}$, instrument-specific bias term $b$, and an ambiguity term $N \cdot \lambda$ in the case of phase observations. The parameter $N$ describes the integer ambiguity of each link, and the parameter $\lambda$ represents the specific wavelength. Table~\ref{tab_simulation_parameters} contains an overview of the essential simulation parameters derived from the observation equations.
\begin{table}[h]
	\caption{Overview of simulation parameters. Table modified from \cite{bib_vol_23}.}\label{tab_simulation_parameters}
		\begin{center}
		\begin{tabularx}{\textwidth}{LLCC}
				\toprule
				\textbf{Parameter} 	& \textbf{Model} 	& \textbf{MWL} 	& \textbf{OPT}\\
				\midrule
				Orbit		 	   	& TLE         		& $\surd$			& $\surd$\\
				Clocks		 	   	& Colored noise		& $\surd$			& $\surd$\\
				TDS offset	 	   	& Inter-technique time bias         		& -			& $\surd$\\
				TDS noise	 	   	& Inter-technique colored noise     		& -			& $\surd$\\
				Local ties  	   	& Inter-technique height offset     & -			& $\surd$\\
				Laser-Reflector-Array  	   	& Champ-Like     & -			& $\surd$\\
				Shapiro 	   		& -         		& $\surd$			& $\surd$\\
				Sagnac		 	   	& 1st \& 2nd order  & $\surd$			& $\surd$\\
				MWL bias 			& Static         	& $\surd$			& -\\
				Ionosphere	 	   	& three-dimensional and time-dependent model    		& $\surd$			& -\\
				Troposphere 	   	& Raytracing   		& $\surd$			& $\surd$\\
				Tropospheric fluctuations 	   	& Time-/Geometry based variance-covariance modelling         		& $\surd$			& $\surd$\\
				\bottomrule
			\end{tabularx}	
		\end{center}
\end{table}
\noindent
See Vollmair et~al. (2023) \cite{bib_vol_23} for a more detailed explanation of the standard simulation and the models used.\\
\noindent
For this study, the existing simulation software was extended to include the following elements:
\begin{itemize}
	\item relativistic red shift,
	\item simulation of more ground stations,
	\item simulation of different (optical) clocks,
	\item time-dependent bias for OPT and MWL.
\end{itemize}
The relativistic red shift must be handled differently for satellite and ground clocks. According to the special theory of relativity, the clock on the satellite experiences time dilation ($\delta \gamma_{\tau}$) due to the higher velocity of the satellite relative to the ground station, as well as an additional gravitational red shift ($\delta \gamma_{z,s}$) according to general relativity theory due to position-dependent gravitational potential. 
Terrestrial time (TT), which refers to the rotating geoid, was chosen as the reference time system for this simulation.
Therefore, for the ground stations the time dilation can be neglected, and only the gravitational red shift ($\delta \gamma_{z,r}$) affecting the ground station clocks. Each of the different atomic clocks at the respective stations has an offset to TT. An earth-fixed/earth-centered coordinate system was chosen as the reference coordinate system.\\
This results in an extension of the clock terms of the observation equations (\ref{eq_theta_1}-\ref{eq_phi_2}), which currently only contain the errors of the different clocks ($\sigma \tau$). The new clock terms can be expressed as follows:

\begin{linenomath}
	\begin{equation}
		\delta \tau^{s}  = \sigma \tau^{s} + \gamma_{\tau,s}+ \gamma_{z,s}		
	\end{equation}
	\begin{equation}
		\delta \tau_{r}  = \sigma \tau_{r} + \gamma_{z,r}
	\end{equation}
\end{linenomath}
The relativistic redshift of the ACES clock was simulated according to Asbhy et al. (2003) \cite{bib_ash_03}. The same model was used for simulation and correction. Based on different orbits used for simulation and processing, an error occurs. Nevertheless, this error is negligible for the evaluation of simultaneously observed passes due to the formation of differences of observations between the groundstations.

To simulate the effect of gravitational redshift on the groundstation clocks Eq.~\eqref{eq_grav_red} is used:
\begin{linenomath}
	\begin{equation}
		z =	\frac{g \cdot H}{c^{2}}\label{eq_grav_red}
	\end{equation}
	with, $g=9.80665$ $\frac{m}{s^{2}}$\\
\end{linenomath}
The parameter $g$ denotes gravity, $H$ denotes the orthometric height of the clock at the respective station, $c$ denotes the speed of light.\\
The ACES clock is an artificial clock generated from the fusion of two clocks. The space hydrogen maser (SHM) is used for short-term stability, and the caesium clock PHARAO is used for long-term stability. Two hypothetical optical clocks are simulated for the ground stations, a transportable optical clock (TOC) and a permanently installed optical clock (OC). Both clocks refer to existing strontium lattice optical clocks. TOC is based on SOC2 \cite{bib_sch_18} and OC is based on \cite{bib_sch_20}. Some ground stations do not have their own optical clock; in the simulation, these were connected to an optical clock at a neighboring ground station via a fiber-optic link. The ELSTAB connection between the PTB in Braunschweig and the GFZ in Potsdam serves as the basis for this \cite{bib_kre_15}. Table~\ref{tab_stations} provides an overview of the various stations, the clocks used in each case, and the measurement techniques available at each site.

\begin{figure}[H]
	\includegraphics[width=13.5 cm]{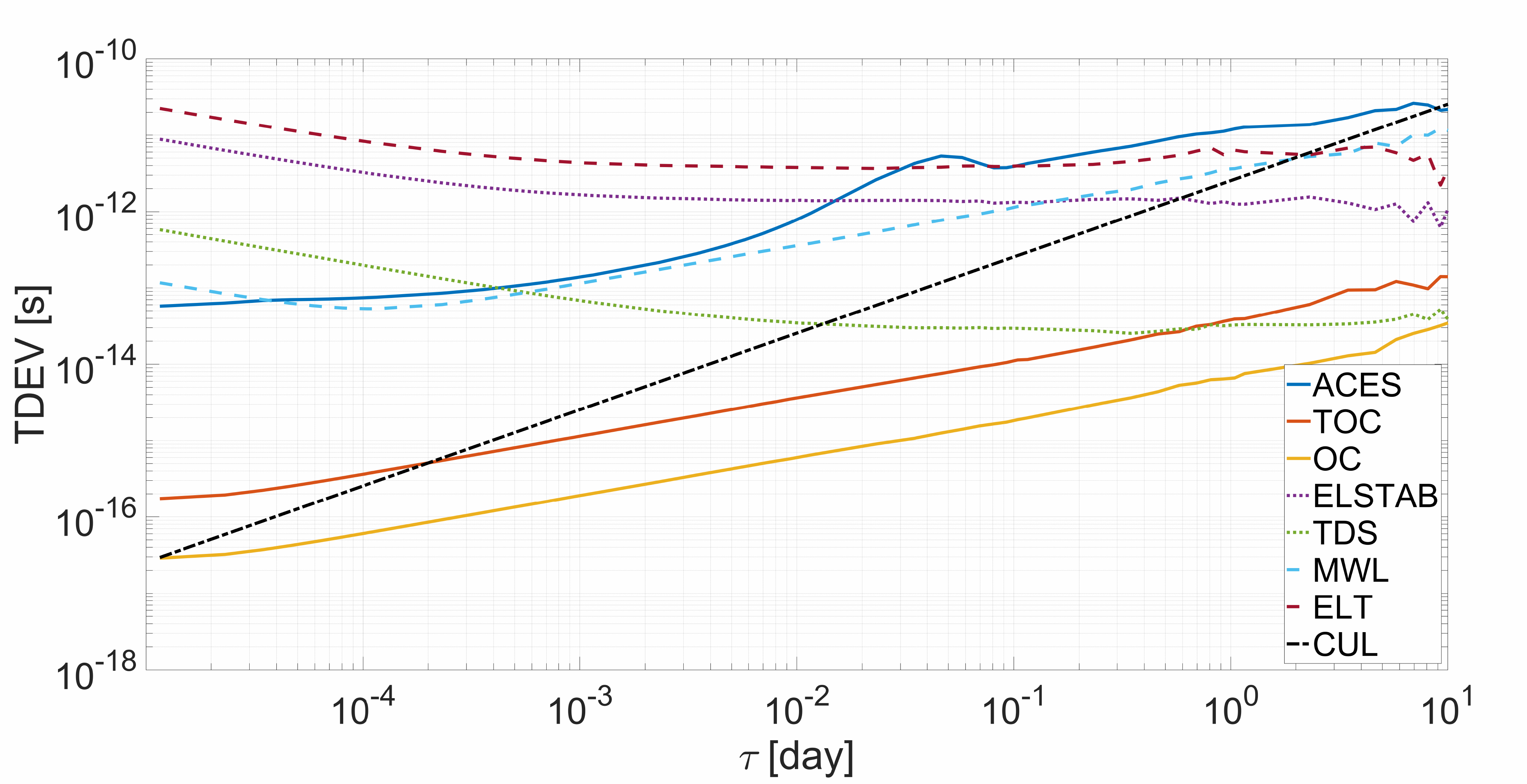}
	\caption{Time deviation of the noise from the simulated clocks and sensors.\label{sensor_tdev}}
\end{figure}

\begin{table}[H] 
	
	\caption{Characterization of the geodetic stations.\label{tab_stations}}
		\begin{tabularx}{\textwidth}{LLLCCCC}
			\toprule
			\textbf{Station}			& \textbf{Clock}					& \textbf{MWL}	& \textbf{ELT}	& \textbf{Height}	& \textbf{Height of connected Clock}\\
			\midrule                	                                                    
			ISS							& ACES								& X				& X				& $~400$ km			& $~400$ km		 \\
			GOW							& TOC			& X				& X				& $ 618.76$ m		& $618.76$ m	 \\
			PTB							& OC				& X				& -				& $  87.33$ m		& $ 87.33$ m	 \\
			SYRTE						& OC				& X				& -				& $  79.54$ m		& $ 79.54$ m	 \\
			GFZ\textsuperscript{1}							& OC\textsuperscript{2}							& -				& X				& $ 103.57$ m		& $ 87.33$ m	 \\
			OCA\textsuperscript{3}							& OC\textsuperscript{2}							& -				& X				& $1268.30$ m		& $ 79.54$ m	 \\
			NERC-SGF\textsuperscript{4}	& OC\textsuperscript{2}								& -				& X				& $  30.96$ m		& $ 42.48$ m	 \\
			BAO \textsuperscript{5}		& OC\textsuperscript{2}								& -				& X				& $  88.96$ m		& $109.28$ m	 \\
			\bottomrule
		\end{tabularx}
	\noindent{\footnotesize{\textsuperscript{1} Geoforschungszentrum (GFZ) connected to Braunschweig (PTB)}}\\	
	\noindent{\footnotesize{\textsuperscript{2} Connected via ELSTAB to OC at different station}}\\		
	\noindent{\footnotesize{\textsuperscript{3} Observatoire Côte d'Azur connected to Paris (SYRTE)}}\\	
	\noindent{\footnotesize{\textsuperscript{4} Natural Environment Research Council-Satellite Geodesy Facility connected to London}}\\	
	\noindent{\footnotesize{\textsuperscript{5} Borowiec Astrogeodynamic Observatory connected to Warsaw}}
	
\end{table}
For stations that do not have an optical clock of their own but do have an ELSTAB connection, the orthometric height at the location of the connected optical clock was used. In this simulation, the PTB/GFZ and SYRTE/OCA stations use the same optical clock, but with an error contribution caused by the ELSTAB system. Only the GOW has a combination of an MWL terminal and an SLR station. In addition to the clock simulation, colored instrument noise were simulated for ELT and MWL.
Table~\ref{tab_noise_sigmas} summarizes the noise characteristics used for the individual sensors.\\
\begin{table}[H] 
	
	\caption{Noise characteristic of simulated clocks and sensors.\label{tab_noise_sigmas}}
		\begin{tabularx}{\textwidth}{LCCCC}
			\toprule
			\textbf{Sensor}				& \textbf{White phase}	& \textbf{Flicker phase}								& \textbf{White frequency}	& \textbf{Flicker frequency}\\
			\midrule                	                                                    
			ACES						& -						& $1\cdot10^{-13}$										& $3\cdot10^{-14}$			& $1\cdot10^{-15}$\\
			TOC							& -						& -														& $3\cdot10^{-16}$			& - \\
			OC							& -						& -														& $5\cdot10^{-17}$			& - \\
			TDS							& $1\cdot10^{-12}$		& $6\cdot10^{-14}$										& -							& - \\
			ELSTAB						& $15\cdot10^{-12}$		& $3\cdot10^{-12}$										& -							& - \\
			MWL							& $2\cdot10^{-13}$		& -														& $3\cdot10^{-14}$			& - \\
			ELT\textsuperscript{1}							& $37\cdot10^{-12}$		& $8\cdot10^{-12}$/$14\cdot10^{-12}$	& -							& - \\
			\bottomrule
		\end{tabularx}
	\noindent{\footnotesize{\textsuperscript{1} Second flicker phase term is used to generate an offset by calculating the mean value of a one day realisation}}\\	

\end{table}
The instrument noise for MWL and ELT consists of white noise and an instrument-specific bias or jitter, which is determined from the colored noise during a pass. For ELT, a daily offset is added in addition to the first continuous realization of the flicker phase noise with a magnitude of $8\cdot10^{-12}$ s. This daily offset is calculated from the arithmetic mean of a daily realization of flicker phase noise with a magnitude of $14\cdot10^{-12}$ s. However, the specified noise of the sensors only describes their instability. For both types of optical clocks, the degree of uncertainty is an important limiting factor in determining the physical height of the clock. According to the specification, the certain uncertainty level (CUL) for the clocks used here is $3\cdot10^{-17}$ s \cite{bib_fal_14}. Using formula \ref{eq_grav_red} results in a maximum height resolution of $27.5$ cm.
Figure~\ref{sensor_tdev} shows the time deviation of the clock errors, the instrument noise for ELT and MWL, and that of the CUL.\\

The observation variables in this study are clock desynchronization ($\Delta t$) calculated from the difference between ground and satellite clocks, which are generated either by OPT or MWL observations.\\
For MWL, differences from pre-corrected phase observations of downlink and uplink in the Ku band of the same epoch are used to determine clock synchronization ($\Delta t_{\phi}$). For this purpose, the ionospheric influence on the signal propagation time and ambiguities are corrected. The ionospheric delay can be minimized by combining the two downlinks in the S and Ku bands. The ambiguities are estimated using a least squares adjustment. The Sagnac effect for downlink and uplink is approximated with sufficient accuracy using an additional orbit model. In addition, the relativistic effects affecting the ACES clock are corrected ($\gamma_{\tau,s} , \gamma_{z,s}$). To calculate clock desynchronization from ELT observations, the optical one-way observations ($\theta_{1}$) must first be assigned to the corresponding SLR observations ($\theta_{2}$). To then calculate $\Delta t$, the difference between the correctly assigned laser-based one-way and two-way measurements can be formed.

\section{Methodology}
\label{sec:Meth}
In the following, we exploit the fact that the initial time offset and slowly varying instrumental biases primarily affect the intercept of $\Delta t(t)$, whereas the gravitational signal manifests as a characteristic rate (slope) over suitable intervals. This motivates the slope-based estimation approach and its application to different visibility configurations as described in the following sections.

\subsection{From Time Differences to Height Differences}


Physical, gravity-related heights can be defined through the Earth’s gravity field by linking geometric position to the gravity potential \cite{bib_Tor_23}. In geodetic practice, the potential difference between two locations 1 and 2 is expressed by the geopotential number
\begin{equation}
C_{P} = -(W_P - W_0) ,
\end{equation}
where $W$ denotes the gravity potential. A height system such as the normal heights relate this potential difference to a height difference by introducing the mean normal gravity. For the normal height difference $\Delta H^N_{12}=H^N_2-H^N_1$, one obtains the approximation
\begin{equation}
\Delta W_{12} \approx \bar{\gamma}\,\Delta H^N_{12},
\end{equation}
where $\bar{\gamma}$ is an appropriately averaged normal gravity value along the normal plumbline. See, e.g., \cite{bib_Tor_23, bib_hei_67, bib_Hof_67} for definitions and conventions.

\begin{figure}[H]
	\includegraphics[width=12 cm]{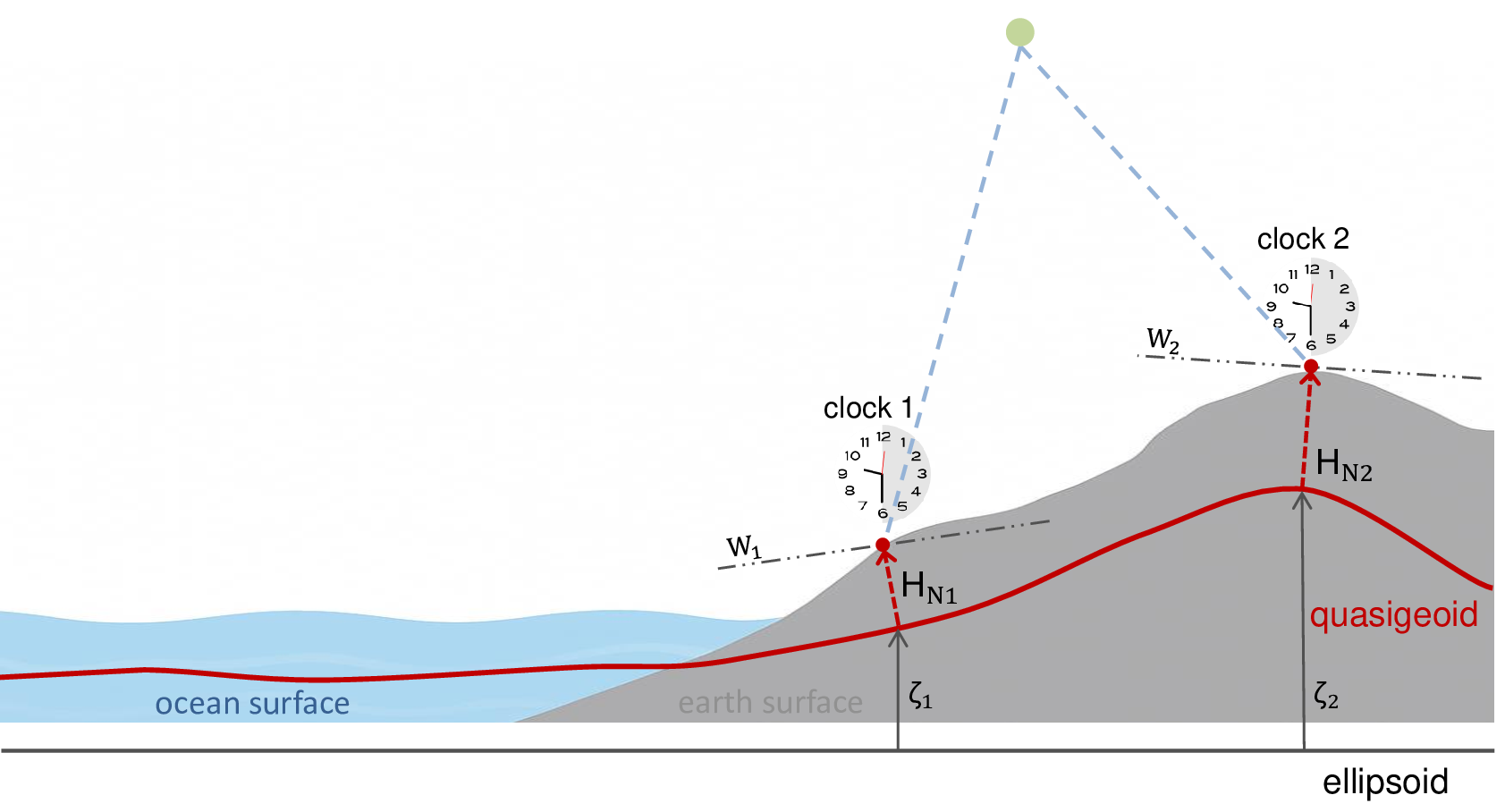}
	\caption{Principle of height determination using the gravitational redshift.}
    \label{KonstHeight}
\end{figure}  

According to general relativity, the rate of an ideal clock depends on the gravity potential. Clocks located at different gravity potentials therefore tick at different rates, an effect known as gravitational redshift. For two stationary clocks, as shown in Figure~\ref{KonstHeight} , the relative time rate difference can be written as
\begin{linenomath}
\begin{equation}
\frac{\Delta \tau_{12}}{\tau_2} \approx \frac{\Delta W_{12}}{c^2} = \frac{\bar \gamma \Delta H_{12}^N}{c^2},
\label{equ:RelTimeDiff}
\end{equation}
\end{linenomath}
where $\Delta\tau_{12}$ denotes the accumulated proper-time difference between the clocks over an interval of duration $\tau_2$, $\Delta W_{12}$ is the gravity potential difference, $c$ is the speed of light, and $\Delta H^N_{12}$ is the corresponding normal height difference. Equation~\eqref{equ:RelTimeDiff} provides the fundamental link between clock comparisons and gravity field related heights \cite{bib_bje_85, bib_meh_18}.

In practical time-transfer experiments, the directly accessible observable is the clock desynchronization or time offset between two sites, thus the difference of clock readings after applying relevant measurement and propagation corrections. In the ACES-related context, clock desynchronization $\Delta t$ is derived from optical (ELT/OPT) and microwave (MWL) links after accounting for relativistic terms, atmospheric propagation effects, and instrumental contributions as described in the simulation framework (see Section \ref{Sim_Back}).
Consequently, the height determination problem can be formulated as estimating $\Delta H^N_{12}$ from a time series of desynchronization observations. A convenient observation model is obtained by rewriting equation \eqref{equ:RelTimeDiff} for the accumulated time offset between two clock sites:
\begin{linenomath}
\begin{equation}
\tau_2 - \tau_1 = \left(H_2^N - H_1^N \frac{\overline{\gamma}_1}{\overline{\gamma}_2}\right)\frac{\overline{\gamma}_2 \, \tau_2}{c^{2}},
\label{equ:Obsv}
\end{equation}
\end{linenomath}
where the $\gamma$-terms represent mean normal gravity values associated with the respective sites and the chosen height system conventions. Equation~\eqref{equ:Obsv} emphasizes that the measurable quantity is a time offset, while the parameter of interest is a gravity-related height difference. Thus, the time offset between two clocks can be formally written as

\begin{equation}
\Delta t(t) = \frac{\bar{\gamma} \Delta H^N_{12}}{c^2} \, t + b ,
\label{equ:Obsv_form}
\end{equation}

where $b$ represents the initial time offset. The height difference is therefore directly proportional to the slope of the time desynchronization.

\subsection{Slope-Based Estimation}
\label{chap:SlopeEst}
As shown in Section 2.1, the physical height difference between two clock sites is directly proportional to the temporal rate difference of their clock desynchronization. In practice, however, the observed time offset $\Delta t(t)$ is affected not only by the gravitational signal but also by the initial time offset, instrumental biases, and stochastic noise contributions.
A direct use of $\Delta t(t)$ would therefore be sensitive to such effects. In contrast, the gravitational redshift results as a linear drift in the desynchronization over sufficiently short observation intervals. This motivates the use of a slope-based estimator. 

In the simulated ACES link data, each satellite pass provides a high-rate set of desynchronization estimates. For the analysis, all observations within one pass are reduced to a single representative value by computing the pass mean. This yields a discrete time series of pass-averaged desynchronization values,
\begin{equation}
\left\{ \big(t_k,\, \overline{\Delta t}_k\big) \right\}_{k=1}^{K},
\end{equation}
where $t_k$ denotes the reference time of pass $k$ (e.g., pass midpoint) and $\overline{\Delta t}_k$ is the corresponding pass mean.

The gravitational signal leads to a constant relative rate difference between the clocks. This implies that the desynchronization evolves linearly in time, with a time derivative given by
\begin{equation}
a_{12} = \frac{d}{dt}\Delta t(t) \approx \frac{\Delta W_{12}}{c^2} = \frac{\bar{\gamma}\,\Delta H_{12}^N}{c^2},
\end{equation}
where $a_{12}$ denotes the relative clock rate difference between the two stations. Therefore, the height difference can be obtained from the slope of a linear model fitted to the pass-mean time series.

At each evaluation epoch $T_m$ (after $m$ available passes), we estimate a single slope parameter $a(T_m)$ by an ordinary least-squares regression using all pass means up to that epoch,
\begin{equation}
\overline{\Delta t}_k = a(T_m)\, t_k + b(T_m) + \varepsilon_k, \qquad k=1,\ldots,m,
\end{equation}
where $b(T_m)$ is corresponding to the initial time offset $b$ in Equation \eqref{equ:Obsv_form} and absorbs bias contributions, while $\varepsilon_k$ represents the residual noise of the pass means. The corresponding height difference estimate is then obtained as
\begin{equation}
\widehat{\Delta H_{12}^N}(T_m) = \frac{c^2}{\bar{\gamma}}\, \widehat{a}(T_m),
\end{equation}

where $\widehat{a}(T_m)$ denotes the estimated slope parameter obtained from the least-squares adjustment, and $\widehat{\Delta H_{12}^N}(T_m)$ indicates estimated quantities.
This cumulative slope estimation is repeated each time a new pass becomes available, resulting in a sequence of progressively refined height estimates. Due to the cumulative nature of the estimation, these values are strongly correlated, and only the final estimate provides the statistically optimal solution. As additional passes are included, the statistical strength of the regression increases. Due to the limited number of passes and the reduction to pass-mean values, the regression remains computationally efficient.
This approach provides two major advantages. First, the initial time offset and slowly varying instrumental biases are largely absorbed by the intercept parameter $b$ and do not affect the slope estimate. Second, the formulation allows a consistent treatment of both common-view and non-common-view observation configurations, as described in the following section.

\subsection{Observation Scenarios}
\label{sec:ObsvScen}

\begin{figure}[H]
    \centering
    \includegraphics[width=8.5cm]{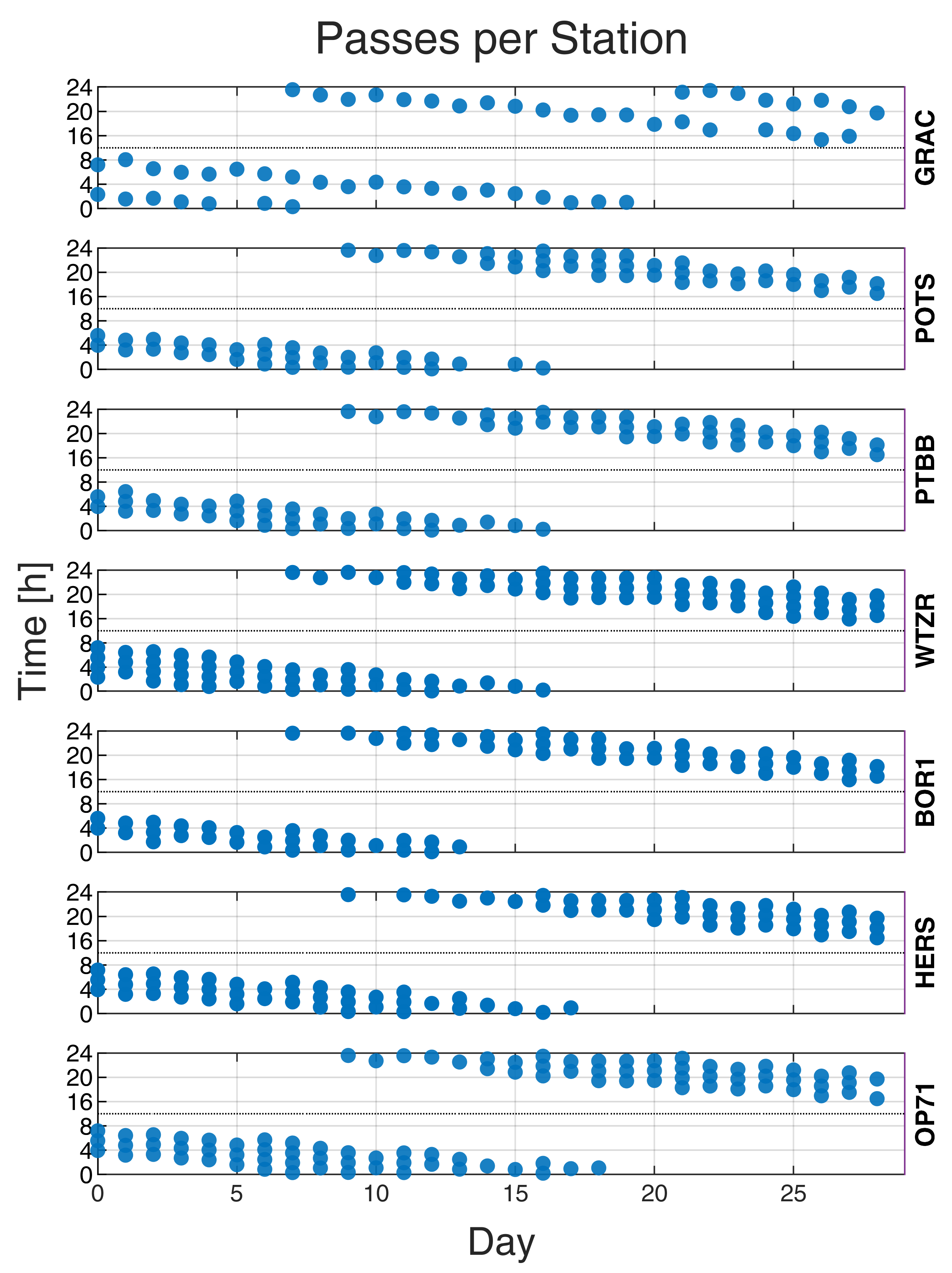}
    \caption{Temporal distribution of ACES satellite passes for all stations considered in this study over the simulation period.}
    \label{fig:passes}
\end{figure}

\begin{figure}[H]
    \centering
    \includegraphics[width=14.5cm]{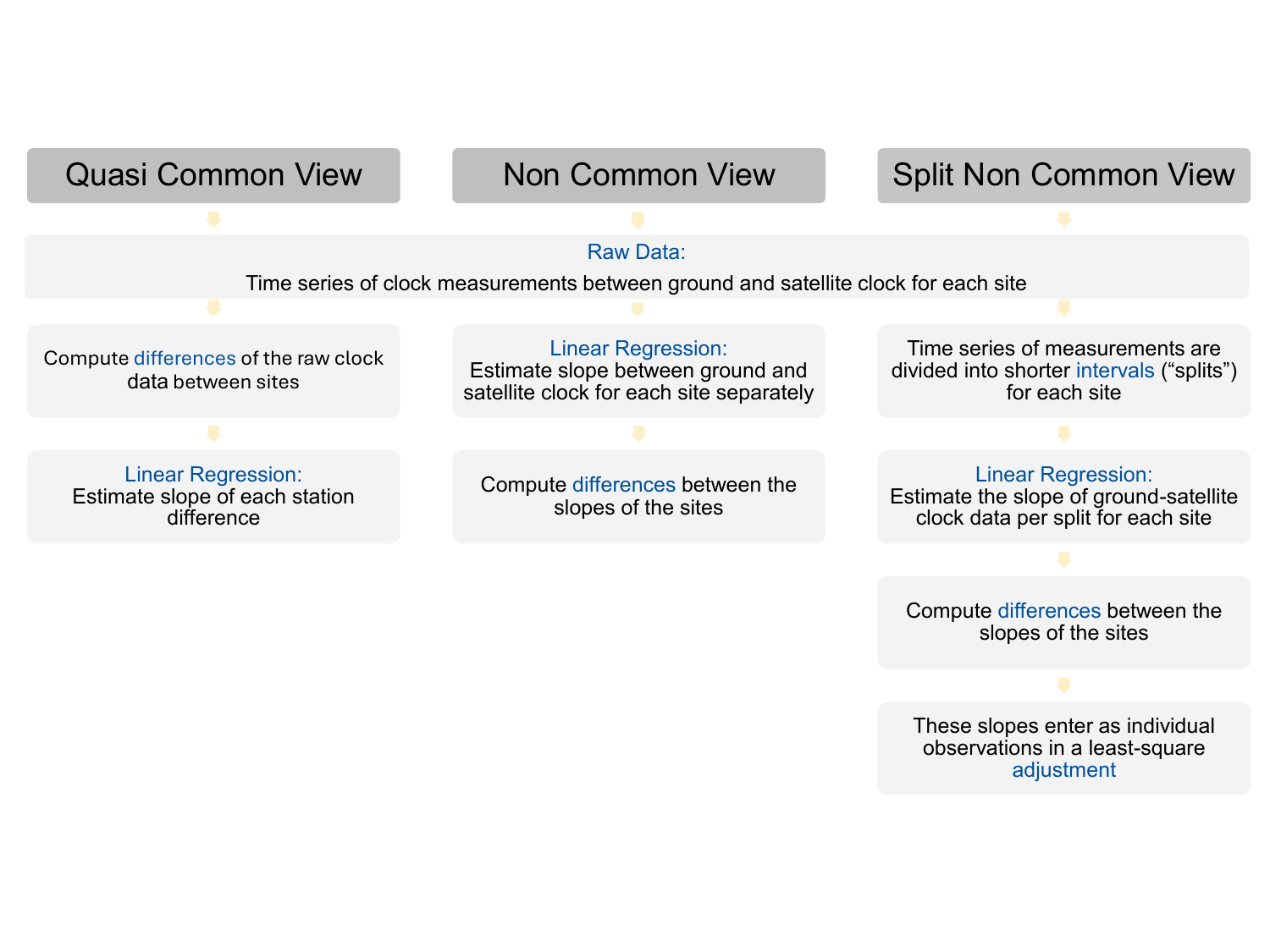}
    \vspace{-60pt}
    \caption{Schematic overview of the processing strategies for the quasi-common view, non-common view, and split non-common view configurations. The figure illustrates the different order of differencing, slope estimation, and adjustment steps.}
    \label{fig:flowchart}
\end{figure}

All observation configurations considered in this study are based on the same underlying data, namely the pass-mean desynchronization time series derived from the ground-satellite clock links. The physical model and the slope-based height determination remain identical for all cases. The ACES orbit geometry results in a limited number of daily satellite passes per station, typically on the order of a few passes per day, each lasting several minutes. Figure~\ref{fig:passes} illustrates the temporal distribution of satellite passes for all stations considered in this study over the simulation period. The figure highlights the irregular spacing of passes and the limited overlap between stations. Simultaneous visibility at two stations, known as common view, is therefore not guaranteed, which motivates the distinction between quasi-common and non-common view configurations. In addition, the high short-term stability of the ACES clock ensemble ensures that clock noise over time intervals of a few minutes is negligible \cite{bib_cac_07, bib_cac_09}. Therefore, allowing a small temporal offset does not significantly degrade the clock comparison, which motivates the usage of quasi- and non-common view constellations. The configurations differ solely in the sequence of processing steps applied to the desynchronization time series. In particular, the order of differencing and slope estimation distinguishes quasi-common view from non-common view approaches, while the split non-common view introduces an additional adjustment process. The different processing strategies are shown in Figure~\ref{fig:flowchart}. The following subsections describe the individual observation scenarios in detail.

\subsubsection{Quasi-Common View}
In the quasi-common view configuration, both stations observe the satellite within a sufficiently short time interval such that the observations can be considered quasi-simultaneous. The duration of this offset is discussed in section \ref{sec:CVOffset}. For each pass, the pass-mean desynchronization values of the two stations are directly differenced, yielding a station-to-station time series,
\begin{equation}
\overline{\Delta t}_{12,k} = 
\overline{\Delta t}_{1,k} - \overline{\Delta t}_{2,k}.
\end{equation}

At a given evaluation epoch, all differenced pass-mean observations collected up to that time are jointly adjusted by linear least squares regression. The estimated slope parameter represents the relative clock rate difference between the two ground stations. Using the relation between clock rate and potential difference given in Equation \ref{equ:RelTimeDiff}, this slope estimate is transformed into the corresponding normal height difference.

\subsubsection{Non-Common View}

In the non-common view configuration, the two stations do not observe the satellite simultaneously. Therefore, station-to-station differences cannot be formed at the observation level. Instead, the slope-based estimation is applied separately to the pass-mean desynchronization time series between each ground and satellite clock.

For each station $i$, all available pass-mean observations up to the considered evaluation epoch are jointly adjusted by linear least squares. This yields a slope estimate $\widehat{a}_i(T_m)$ representing the relative rate difference between the ground clock and the satellite clock. The relative clock rate difference between the two stations is then obtained by differencing the station-wise slope estimates,
\begin{equation}
\widehat{a}_{12}(T_m) = 
\widehat{a}_1(T_m) - \widehat{a}_2(T_m).
\end{equation}

In contrast to the quasi-common view, the regression step precedes the differencing. This enables the determination of the height difference independently of simultaneous satellite visibility at the two sites.

\subsubsection{Split Non-Common View}

The split non-common view extends the non-common view configuration by introducing an additional temporal subdivision of the desynchronization time series. Instead of performing a single regression using all available pass means per station, the data are divided into shorter non-overlapping time intervals.

For each interval $j$ and station $i$, a local slope estimate $\widehat{a}_{i,j}$ is obtained by linear regression. Station-to-station slope differences are then formed for each interval,
\begin{equation}
\widehat{a}_{12,j} = 
\widehat{a}_{1,j} - \widehat{a}_{2,j}.
\end{equation}

The resulting set of slope differences is treated as a collection of observations in a subsequent least-squares adjustment. In principle, the interval-wise slope estimates are not statistically independent, as they are derived from time series that may include temporal correlations and shared error sources. 
For simplicity, these correlations are neglected in the present study and the observations are treated as uncorrelated with equal weighting. Consequently, the weight matrix $\mathbf{P}$ is chosen as the identity matrix. A more realistic treatment including the full covariance structure is left for future work.
With the estimation a single height difference parameter is estimated from multiple interval-wise slope estimates.
By introducing an additional estimation level, the split non-common view enables a more flexible combination of temporally distributed information and separates local slope estimation from the final parameter adjustment.

\section{Results}
\label{sec:results}
This section presents the results of the simulation-based analysis of clock-based height determination using satellite time-transfer observations. The performance of the slope-based estimation approach is evaluated under different observation configurations, including quasi-common view, non-common view, and split non-common view scenarios. 
The analysis focuses on the temporal evolution of the height estimation errors and the impact of the different time-transfer techniques and observation geometries.

\subsection{Observation Setup}
\begin{figure}[H]
	\centering
	\includegraphics[width=11 cm]{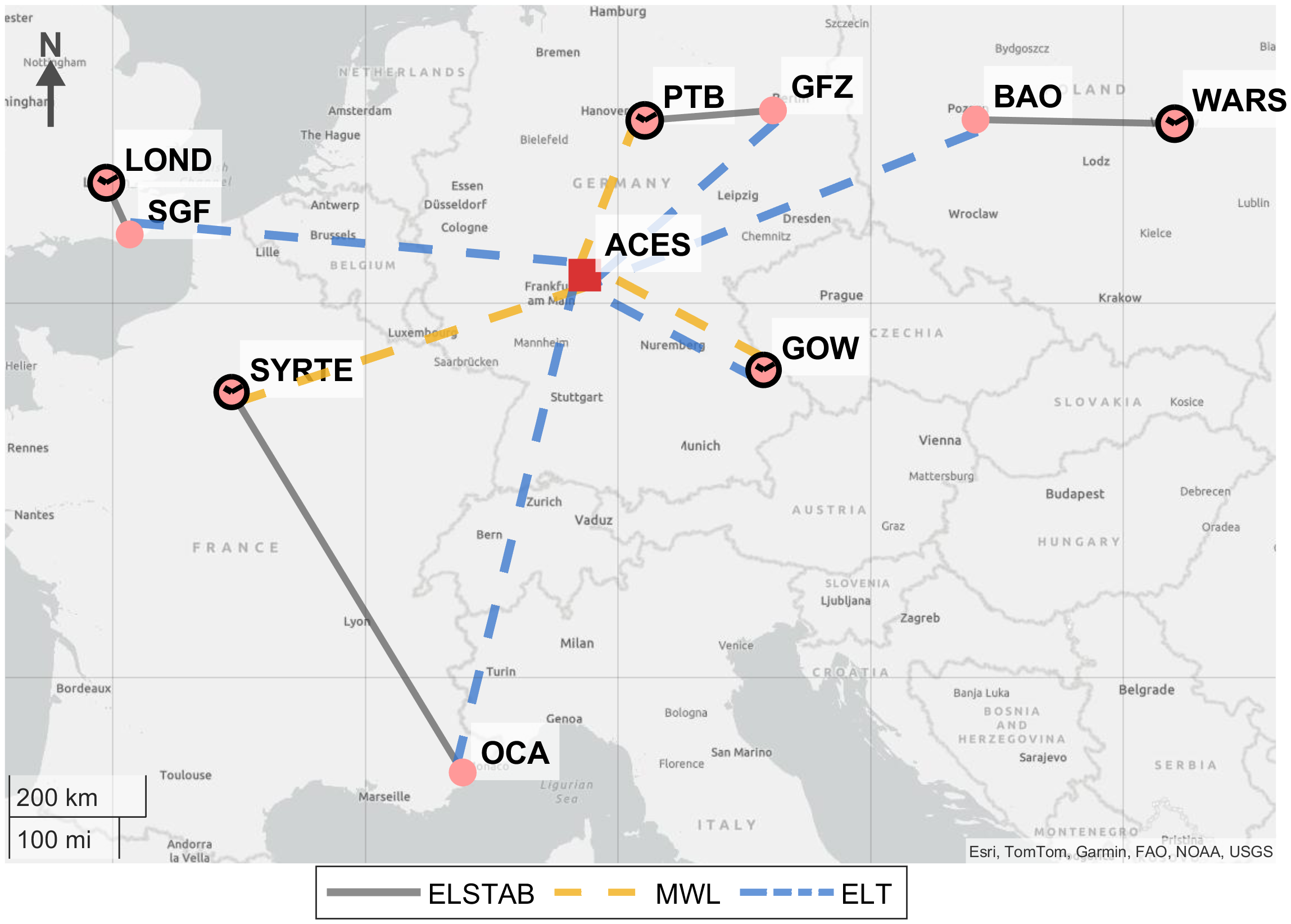}
	\caption{Configuration of the simulated ACES observation network. 
	Ground stations are connected to the satellite via microwave and/ or optical links. 
	Some stations share a common optical clock via stabilized fiber connections.
    Station abbreviations: SYRTE (Paris Observatory), GOW (Geodetic Observatory Wettzell), PTB (Physikalisch-Technische Bundesanstalt, Braunschweig), OCA (Observatoire de la Côte d’Azur), GFZ (German Research Centre for Geosciences, Potsdam), BAO (Borowiec Astrogeodynamic Observatory), SGF (NERC Satellite Geodesy Facility).}
	\label{fig:network_map}
\end{figure}

The considered observation network is illustrated in Figure~\ref{fig:network_map}. 
It consists of several European ground stations equipped with either microwave links, optical links via the European Laser Timing experiment, or both, together with optical clock locations connected through stabilized fiber links.

The ACES payload onboard the International Space Station provides the space reference clock, which is compared with ground clocks during satellite passes. Not all stations host their own optical clock. In several cases, a station is connected via a stabilized optical fiber link (e.g., an ELSTAB link \cite{bib_kre_15}) to an optical clock located at a different site. Such links enable the transmission of highly stable frequency or time signals over long distances with negligible degradation of the frequency stability \cite{bib_kre_15, bib_pre_12}. Similar systems exist in other countries and are widely used in optical clock networks.
Consequently, the physical height associated with a station pair always refers to the height of the clock locations rather than to the antenna or telescope positions of the observing stations. In contrast, satellite visibility and thus the availability of observations depend on the geographical position of the measurement stations. As a result, the effective observation geometry is determined by the ground stations, while the gravity-related height difference to be estimated corresponds to the remote clock sites connected to those stations.

Observation epochs are defined by the availability of satellite passes. The reference time associated with each pass corresponds to the pass midpoint. All configurations considered in this study use simulated satellite passes over Europe during a 30-day period. In the quasi-common-view configuration,  only satellite passes that are observed within a short time offset at both stations are used, shown in Section \ref{sec:CVOffset}. Whereas the non-common-view configurations exploit all available passes independently for each station. 

To evaluate the results, height estimation errors are computed as the difference between the estimated normal height difference and the simulated reference value. The error evolution is analyzed as a function of observation time for QCV and NCV or interval length for Split NCV.
The analysis assumes that the ground clocks operate continuously without interruptions over the considered time span. Under this assumption, the desynchronization between clocks evolves smoothly in time and can be modeled by linear segments corresponding to the gravitational redshift signal. In practice, real clocks may experience interruptions or frequency adjustments, which would require restarting the slope estimation. These effects lie beyond the scope of the present study and will be the subject of future investigations and publications.

\subsubsection{Quasi-Common View Offset}
\label{sec:CVOffset}
\begin{table}[H] 
	\caption{Number of quasi-common view observations for different station pairs and link types. Ground station abbreviations: GOW (Wettzell), PTB (Braunschweig), SYRTE (Paris), OCA (Nice), GFZ (Potsdam), SGF (Herstmonceux), BAO (Borowiec).\label{tab:CV_Offset}}	
	\begin{tabularx}{\textwidth}{L L C C C C}
		\toprule
		\multirow{2}{*}{\textbf{Link}} & 
		\multirow{2}{*}{\textbf{Station pair}} & 
		\multicolumn{4}{c}{\textbf{Maximum allowed time offset}}\\
		\cmidrule(lr){3-6}
		& & 0 min & $\leq$ 2 min & $\leq$ 5 min & $\leq$ 15 min\\
		\midrule
		
		\multirow{3}{*}{MWL} 
		& SYRTE--GOW & 0 & 82 & 82 & 82 \\
		& SYRTE--PTB & 0 & 71 & 71 & 71 \\
		& GOW--PTB & 5 & 73 & 73 & 73 \\
		
		\midrule
		
		\multirow{10}{*}{ELT}
		& OCA--GOW & 0 & 29 & 29 & 29 \\
		& OCA--GFZ & 0 & 9 & 9 & 9 \\
		& GOW--GFZ & 17 & 70 & 70 & 70 \\
        & OCA--BAO & 0 & 0 & 15 & 15 \\
		& BAO--GOW & 0 & 71 & 71 & 71 \\
		& BAO--GFZ & 0 & 65 & 65 & 65 \\
        & GFZ--SGF & 0 & 10 & 60 & 60 \\
		& SGF--OCA & 0 & 14 & 15 & 15 \\
		& GOW--SGF & 0 & 15 & 74 & 74 \\
        & SGF--BAO & 0 & 0 & 55 & 55 \\
		
		\bottomrule
	\end{tabularx}
\end{table}
Table~\ref{tab:CV_Offset} summarizes the number of quasi-common view observations for the considered station pairs and link types under different constraints on the maximum allowed time offset between the corresponding satellite passes. For most station pairs, no strictly simultaneous observations are available, which is shown by the absence of common views for a zero-minute offset. Allowing small temporal offsets significantly increases the number of usable observations. In particular, permitting a maximum offset of two minutes already yields the full set of quasi-common view passes for most station pairs using the MWL link. Increasing the threshold further to five or fifteen minutes does not lead to a noticeable increase in the number of observations for these links. A similar behavior is observed for the ELT links. For several station pairs, however, the number of quasi-common view observations increases when the allowed offset is relaxed from two to five minutes, indicating a larger temporal separation of the corresponding visibility windows. Beyond five minutes, the number of available quasi-common views remains unchanged for all station pairs considered. This behavior is mainly driven by the along-track motion of the ACES satellite and the resulting time shift between the visibility windows at different ground stations. Following the assumption in Section \ref{sec:ObsvScen}, a small temporal offset does not significantly degrade the clock comparison. Thus, a maximum allowed time offset of five minutes is used to define quasi-common view observations.

\subsection{Quasi-Common View}
\begin{figure}[H]
	\centering
	
	\begin{minipage}{0.49\textwidth}
		\centering
		\includegraphics[width=\linewidth]{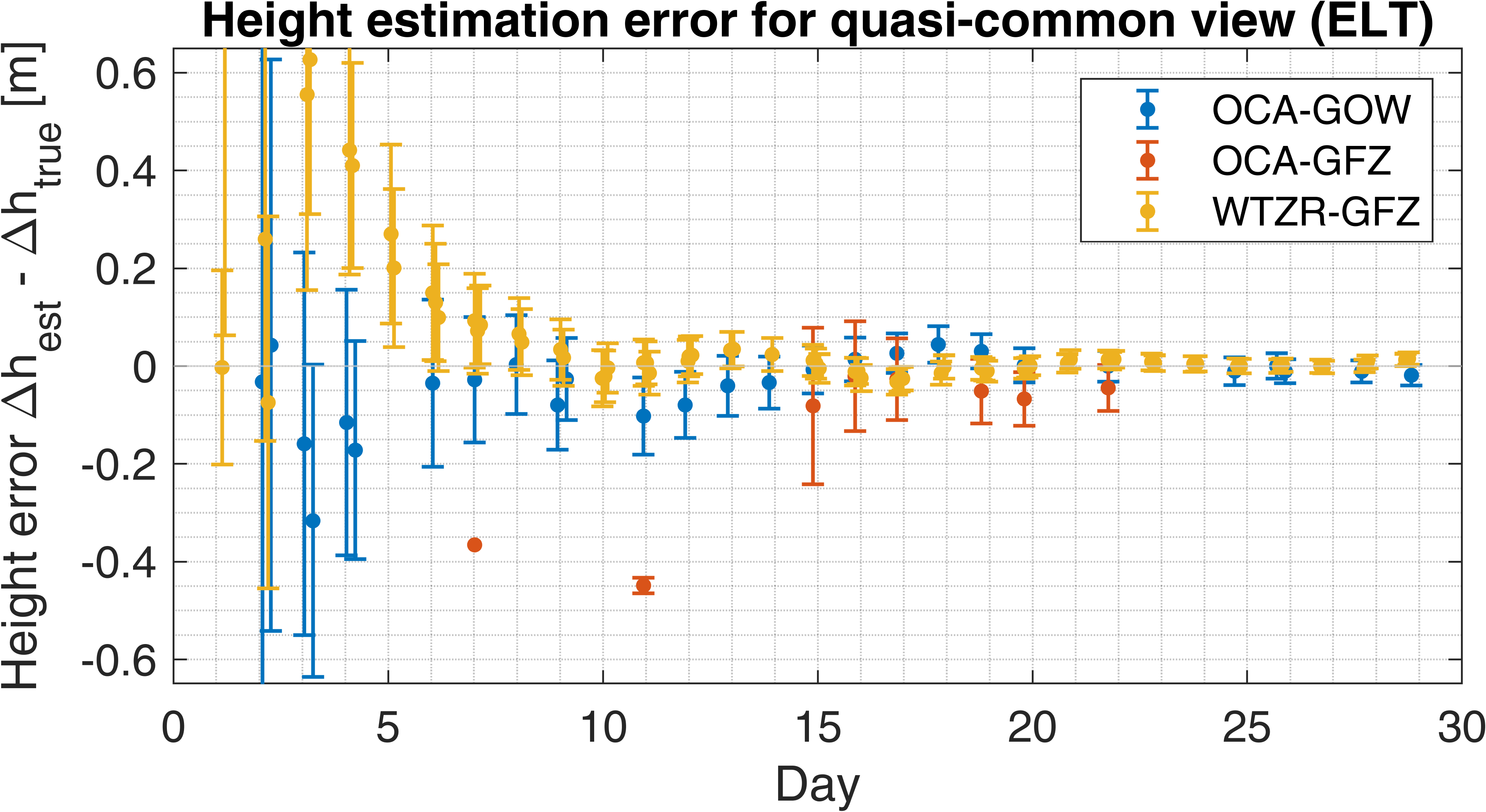}
        \vspace{-20pt}
		\caption*{(a) ELT links (3 station pairs)}
        \vspace{5pt}
	\end{minipage}
	\hfill
	\begin{minipage}{0.49\textwidth}
		\centering
		\includegraphics[width=\linewidth]{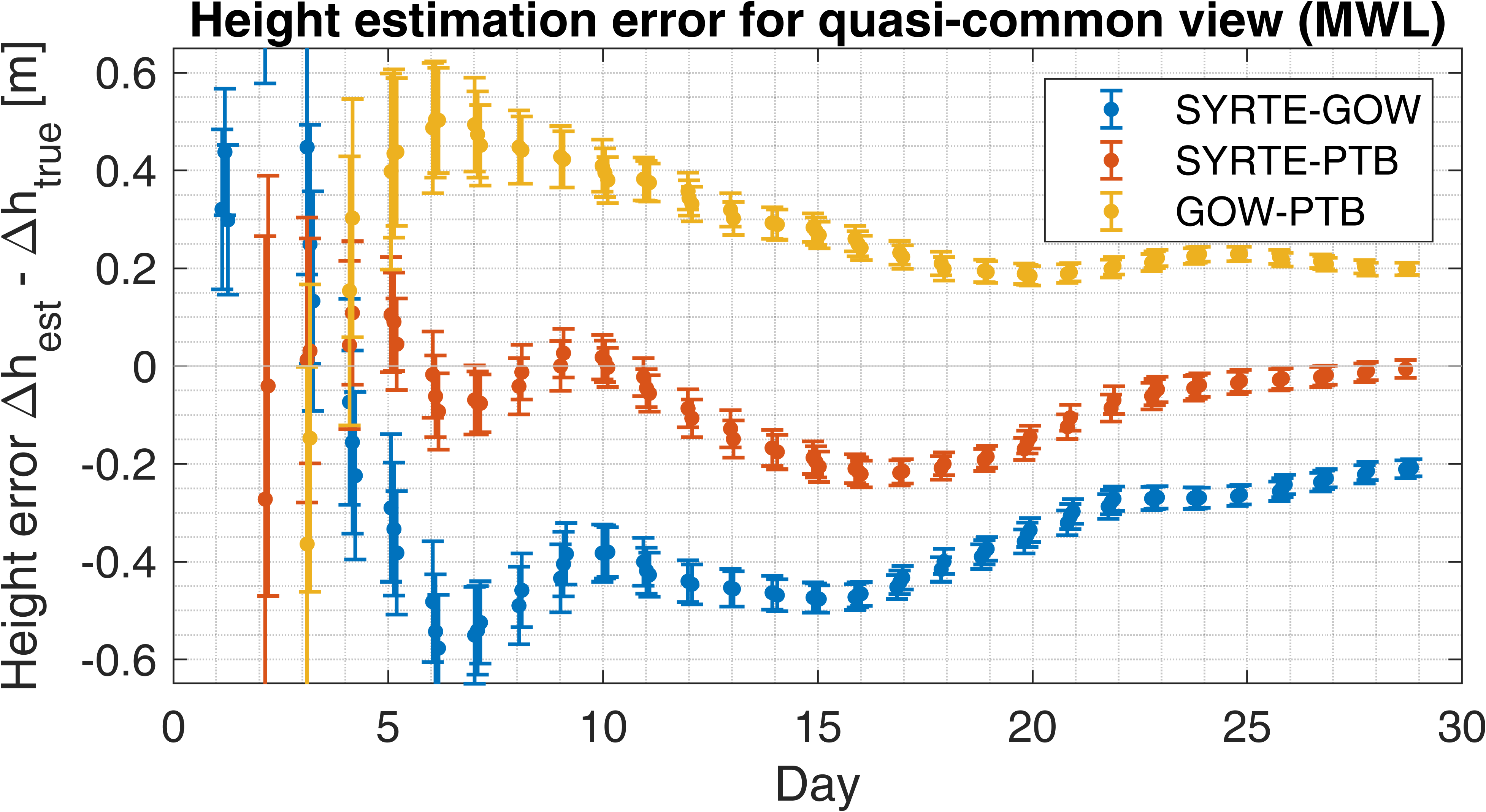}
        \vspace{-20pt}
		\caption*{(b) MWL links}
        \vspace{5pt}
	\end{minipage}
	
	\vspace{0.8em}
	
	\begin{minipage}{0.647\textwidth}
		\centering
		
		\begin{minipage}{0.75\linewidth}
			\includegraphics[width=\linewidth]{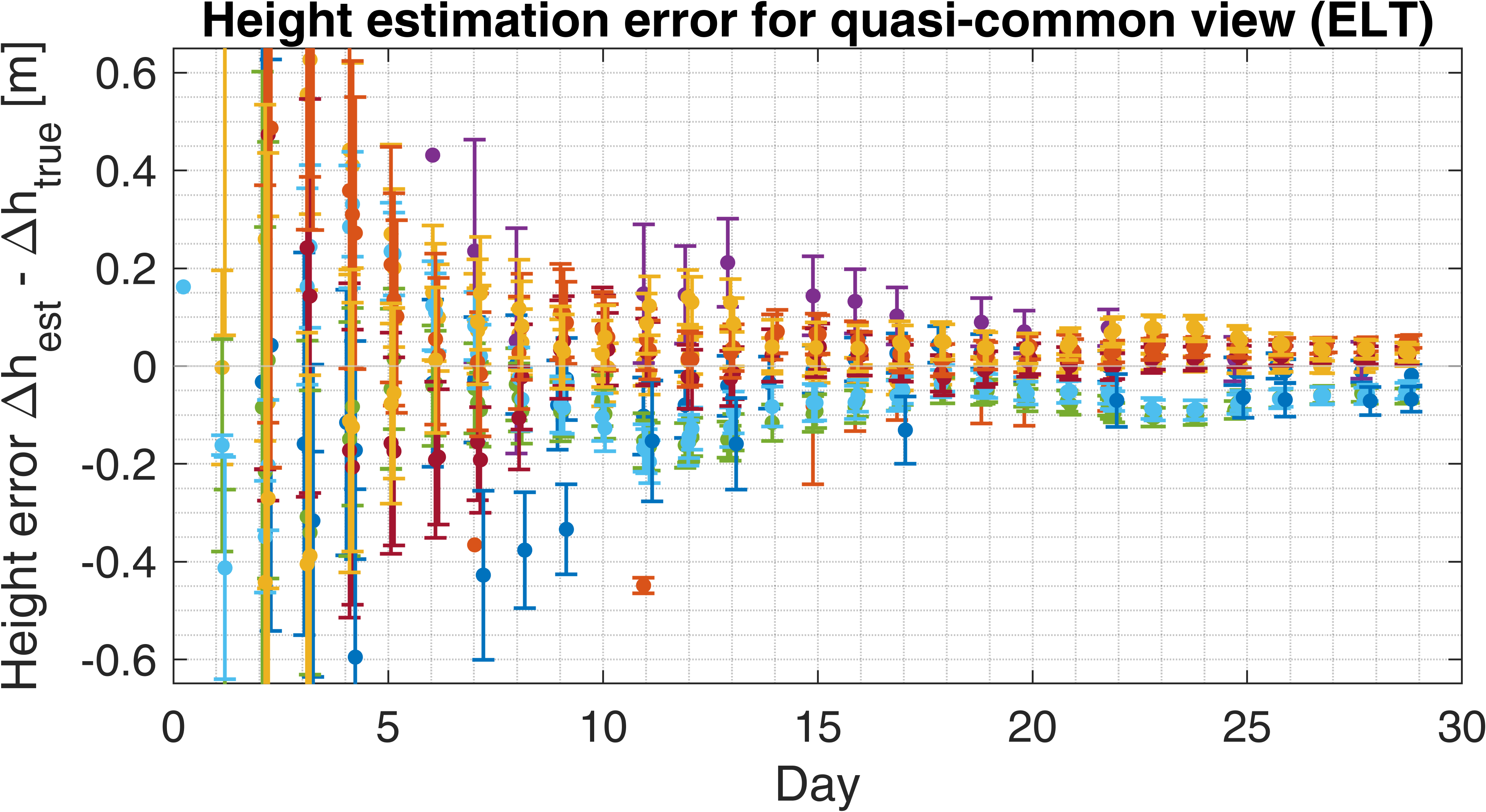}
            \vspace{-20pt}
			\caption*{(c) ELT links (all station pairs)}
            \vspace{5pt}
		\end{minipage}
		\begin{minipage}{0.2\linewidth}
			\includegraphics[width=\linewidth]{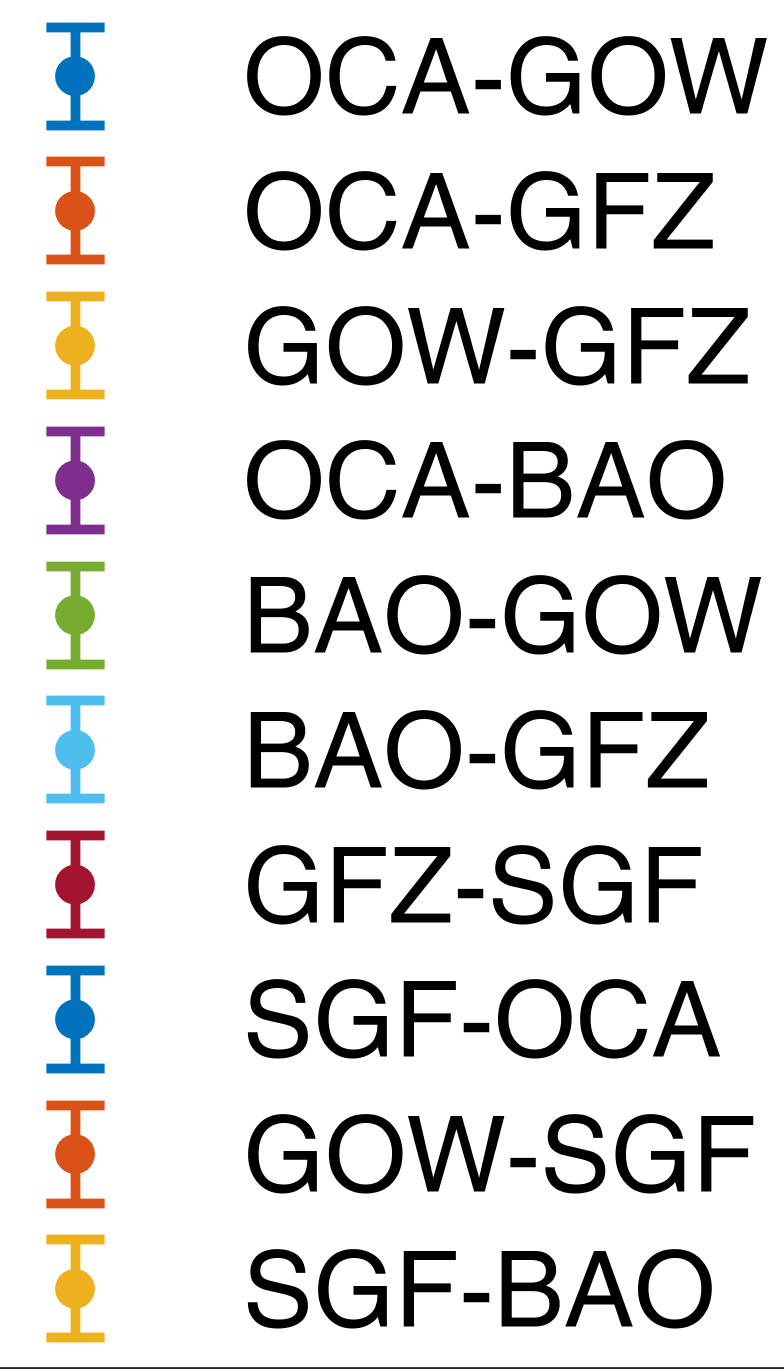}
            \vspace{10pt}
		\end{minipage}
		
	\end{minipage}
	\hfill
	\begin{minipage}{0.48\textwidth}
	\end{minipage}
	
	\caption{Height estimation error for quasi-common view observations. (a) shows ELT results for three selected station pairs, (b) shows all available MWL station pairs, 
and (c) shows the complete set of ELT station pairs.}
	
	\label{fig:QCV_combined}
\end{figure}

The height estimation errors obtained from quasi-common view observations are shown in Figure~\ref{fig:QCV_combined}. For the ELT links, a clear improvement of the height estimates with increasing observation time can be observed. In particular, the station pairs connected to Wettzell reach height errors below $20~$cm after about two to three days. After approximately one week, the corresponding solutions stabilize and remain at the decimeter to few-centimeter level for the rest of the simulation period.
In contrast, the MWL solutions show a less regular behavior. Although all considered station pairs reach height errors of about $20~cm$ after roughly three days, the estimates subsequently exhibit larger fluctuations and, in some cases, a noticeable degradation. This indicates that the MWL-based solutions are influenced by additional error contributions that do not average down like purely white noise.
This difference is consistent with the simulation model for the MWL, where the observation equations include instrument-specific bias terms, ambiguity-related parameters, and ionospheric corrections, while the simulated sensor noise contains not only white phase noise but also white frequency noise and pass-dependent jitter contributions. Such temporally correlated and bias-like error components reduce the long-term stability of the MWL-based slope estimates. For ELT, on the other hand, the slope-based formulation appears to suppress a substantial part of the bias contributions, such that the expected averaging behavior is more clearly visible.
Overall, the quasi-common view results indicate that optical links provide faster convergence and higher long-term stability than microwave links. However, quasi-common view observations require overlapping satellite visibility at both stations. Since such visibility cannot always be guaranteed for all station pairs, e.g. ELT connection OCA-PTB in Figure~\ref{fig:QCV_combined}(a), the following section investigates the corresponding non-common-view solutions.

\subsection{Non-Common View}
Table~\ref{tab:passes_per_station} summarizes the cumulative number of available satellite passes for the individual stations. In contrast to the quasi-common view case, the non-common view configuration is not restricted by simultaneous satellite visibility at two sites. As a result, the number of usable observations increases, particularly for longer observation periods. Already after one to three days, several passes are available for all stations, while after 30 days the total number ranges from 56 to 97 for the ELT stations and from 73 to 97 for the MWL stations.
\begin{table}[H] 
	\caption{Number of satellite passes per station for different observation durations in the NCV case.\label{tab:passes_per_station}}	
	
	\begin{tabularx}{\textwidth}{L L C C C C}
		\toprule
		\multirow{2}{*}{\textbf{Link}} & 
		\multirow{2}{*}{\textbf{Station}} & 
		\multicolumn{4}{c}{\textbf{Number of passes after}}\\
		\cmidrule(lr){3-6}
		& & 1 day & 2 days & 3 days & 30 days\\
		\midrule
		
		\multirow{3}{*}{MWL} 
		& SYRTE & 2 & 5 & 8 & 94 \\
		& GOW   & 3 & 6 & 10 & 97 \\
		& PTB   & 1 & 4 & 6 & 73 \\
		
		\midrule
		
		\multirow{5}{*}{ELT}
		& OCA & 1 & 3 & 5 & 56 \\
		& GFZ & 1 & 3 & 5 & 70 \\
		& BAO & 1 & 3 & 6 & 71 \\
		& SGF & 2 & 5 & 8 & 82 \\
		& GOW & 3 & 6 & 10 & 97 \\
		
		\bottomrule
	\end{tabularx}
\end{table}

\begin{figure}[H]
	\centering
	
	\begin{minipage}{0.49\textwidth}
		\centering
		\includegraphics[width=\linewidth]{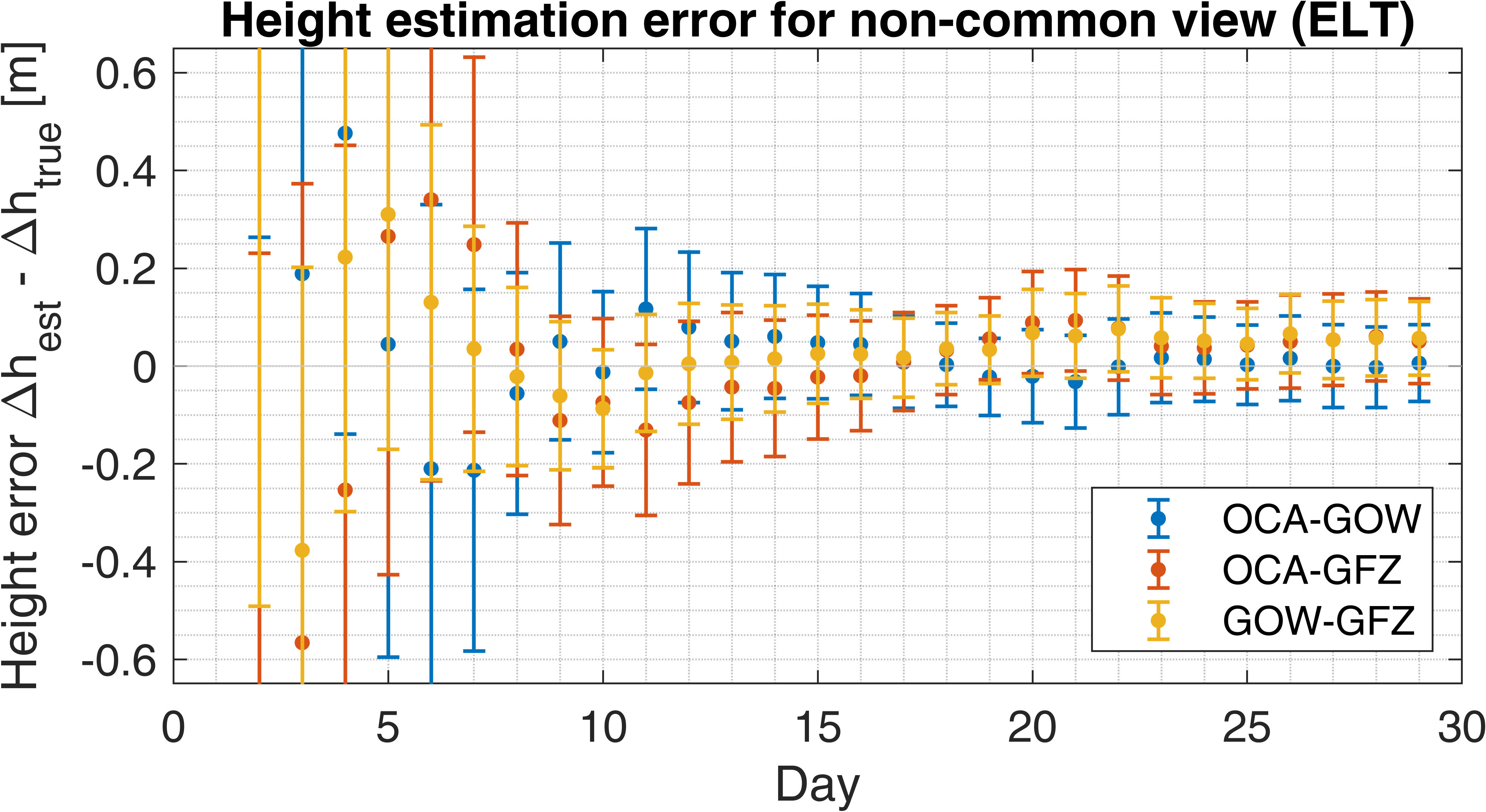}
        \vspace{-20pt}
		\caption*{(a) ELT links (3 station pairs)}
        \vspace{5pt}
	\end{minipage}
	\hfill
	\begin{minipage}{0.49\textwidth}
		\centering
		\includegraphics[width=\linewidth]{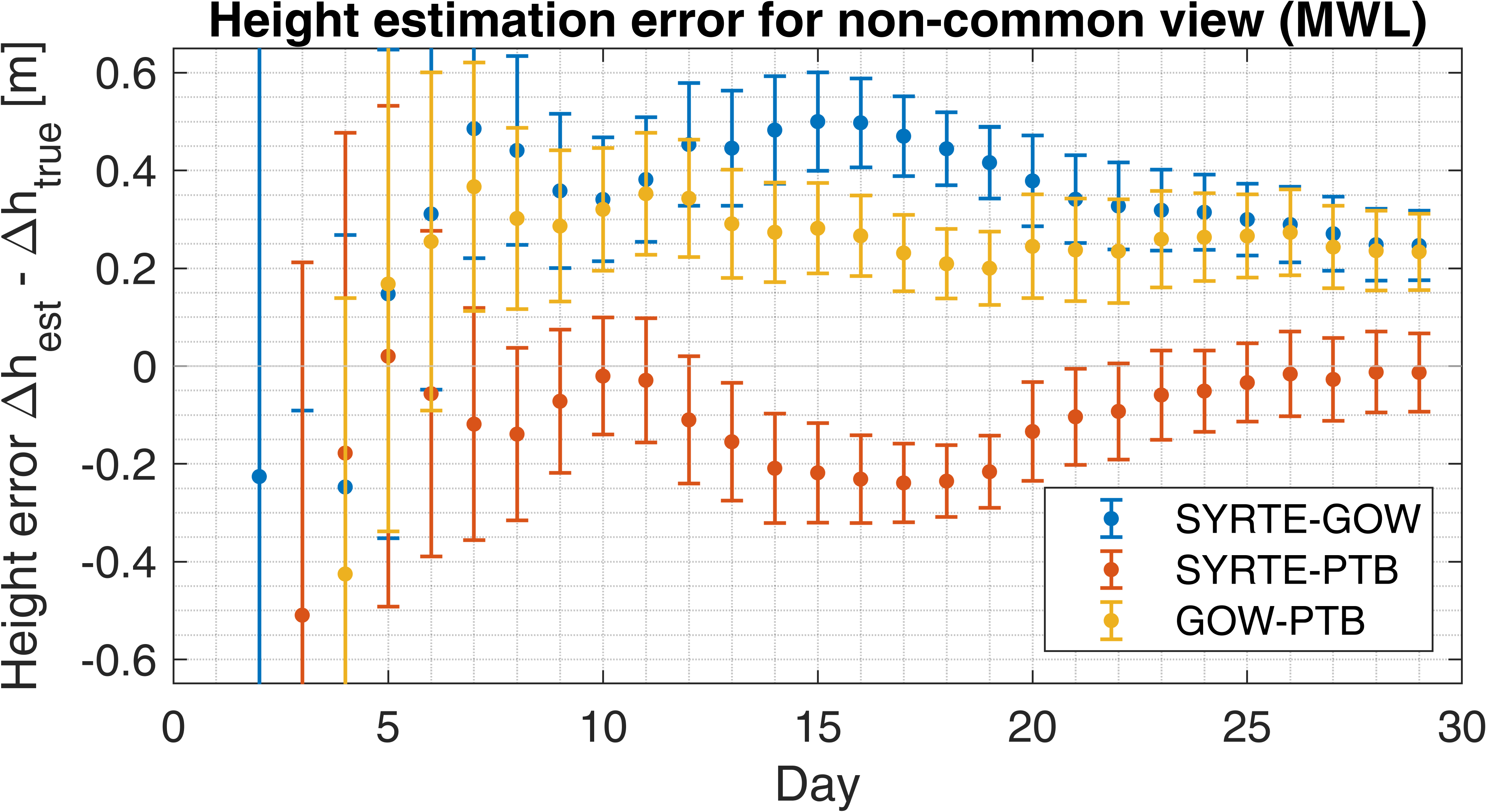}
        \vspace{-20pt}
		\caption*{(b) MWL links}
        \vspace{5pt}
	\end{minipage}
	
	\vspace{0.8em}
	
	\begin{minipage}{0.647\textwidth}
		\centering
		
		\begin{minipage}{0.75\linewidth}
			\includegraphics[width=\linewidth]{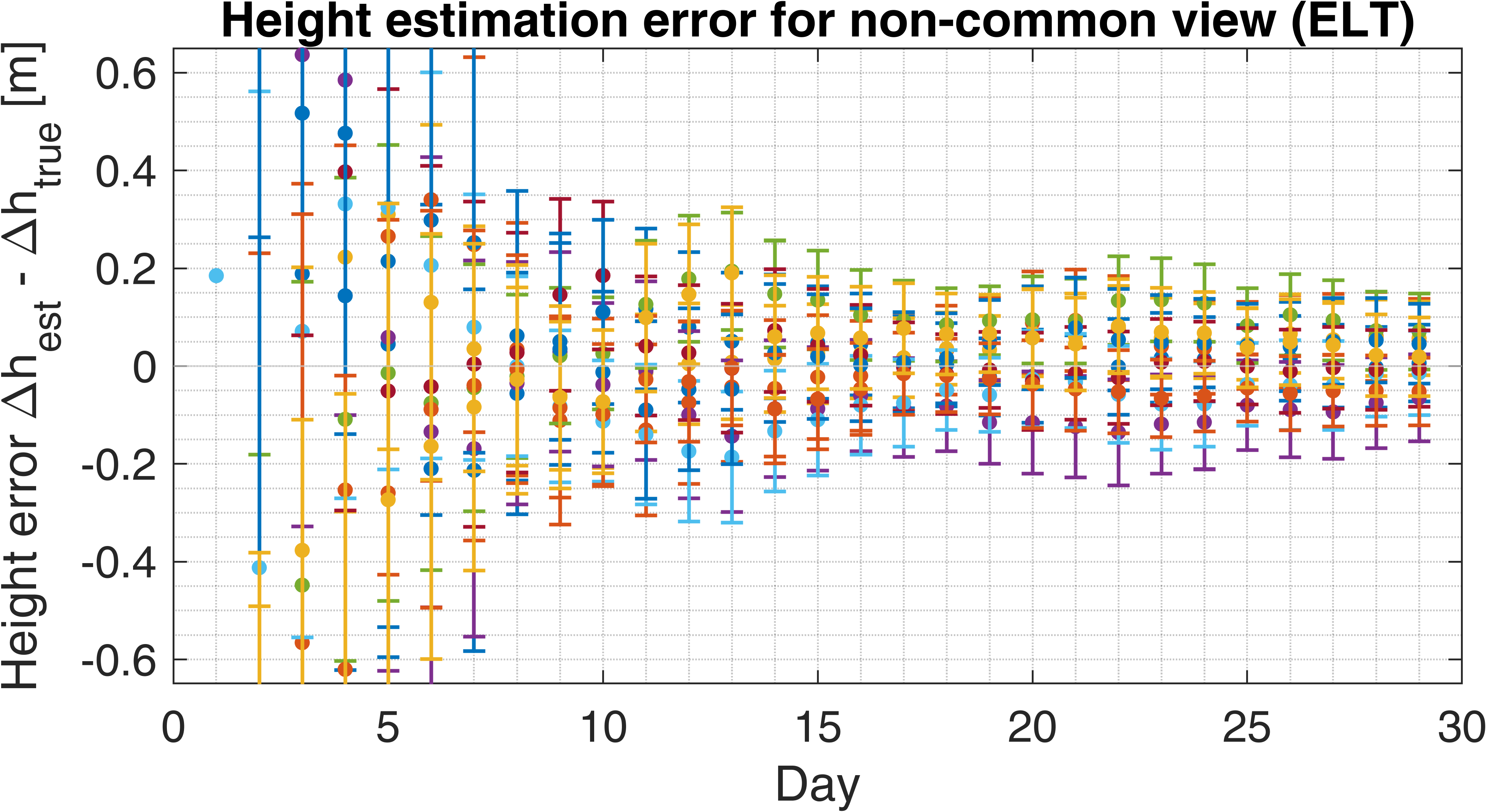}
            \vspace{-20pt}
			\caption*{(c) ELT links (all station pairs)}
            \vspace{5pt}
		\end{minipage}
		\begin{minipage}{0.2\linewidth}
			\includegraphics[width=\linewidth]{figures/legend_all_ELT.png}
            \vspace{10pt}
		\end{minipage}
		
	\end{minipage}
	\hfill
	\begin{minipage}{0.48\textwidth}
	\end{minipage}
	
	\caption{Height estimation error for non-common view observations. (a) shows ELT results for three selected station pairs, (b) shows all available MWL station pairs, and (c) shows the complete set of ELT station pairs.}
	
	\label{fig:NCV_combined}
\end{figure}

The resulting height estimation errors are shown in Figure~\ref{fig:NCV_combined}. Compared to the quasi-common view solutions, the non-common view approach provides a denser observation basis and therefore a more continuous development of the estimated height differences.
For the ELT links, the height errors decrease more regularly with increasing observation time than for the MWL links. After the initial observation period, most ELT-based station pairs converge towards the zero line and remain within the decimeter range, while several solutions approach the few-centimeter level. This behavior is also visible in the full network shown in Figure~\ref{fig:NCV_combined}(c), where the majority of station pairs cluster increasingly close to zero as more passes are included.
The MWL results show a less favorable behavior. Although the larger number of usable passes improves the continuity of the solutions, noticeable systematic deviations remain for all three considered station pairs. In particular, the OP71--WTZR and WTZR--PTBB links remain clearly biased over the full observation period, while only the OP71--PTBB pair approaches the zero line more closely. This is particularly noteworthy because the three MWL station pairs already provided a comparatively large number of observations in the quasi-common view case. Thus, the increased observation availability in non-common view does not fully compensate for the larger and more temporally correlated error contributions affecting the microwave link.

The difference between ELT and MWL is again consistent with the simulated observation model in the MWL case, where noise contributions affect the stability of the slope estimates. In contrast, the optical link solutions benefit from the higher short-term stability of the link and from the reduced sensitivity of the slope-based estimator to offset-like errors.
Overall, the non-common view results demonstrate that the method substantially relaxes the geometrical constraints of quasi-common view processing and is therefore better suited for practical ACES applications. In particular, the combination of non-common view processing with optical links provides a robust basis for the determination of height differences even in the absence of simultaneous satellite visibility.
Compared to the quasi-common view solutions, the general behavior of the individual links remains largely unchanged, although the errors are typically larger by a few centimeters. This indicates that additional error sources associated with non-common view processing, such as satellite clock errors or uncorrelated propagation effects, do not dominate the solution. Instead, the performance continues to be primarily governed by the characteristics of the respective time-transfer links.

To further exploit the increased number of available observations, the following section investigates the split non-common view approach.

\subsection{Split Non-Common View}
\begin{figure}[H]
	\centering
	
	\begin{minipage}{0.75\textwidth}
		\centering
		
		\begin{minipage}{0.78\linewidth}
			\centering
            \hspace*{-0.275\linewidth}
			\includegraphics[width=\linewidth]{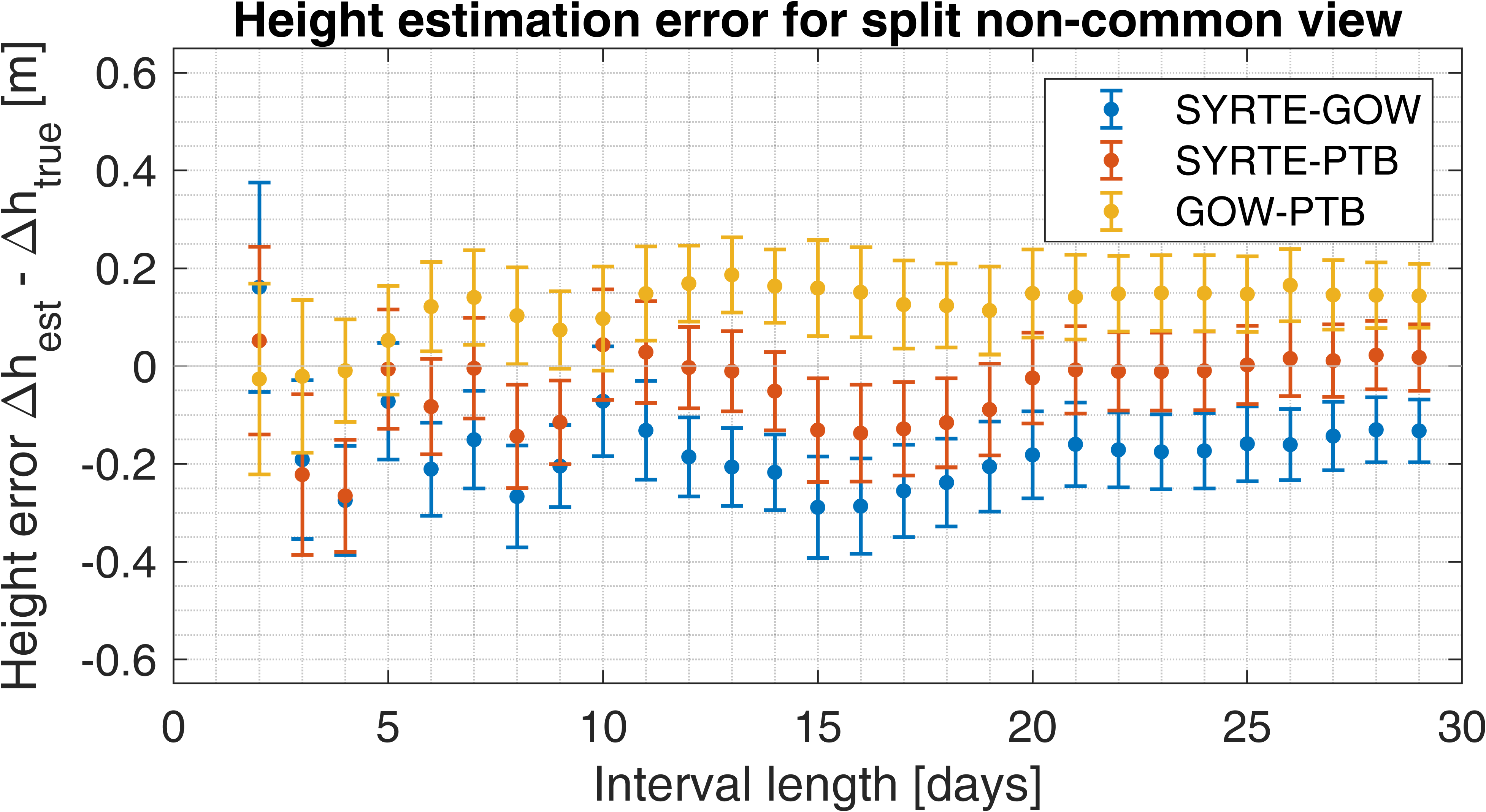}
			\caption*{(a) Three selected station pairs}
		\end{minipage}
		
		\vspace{1em}
		
		\begin{minipage}{0.78\linewidth}
			\includegraphics[width=\linewidth]{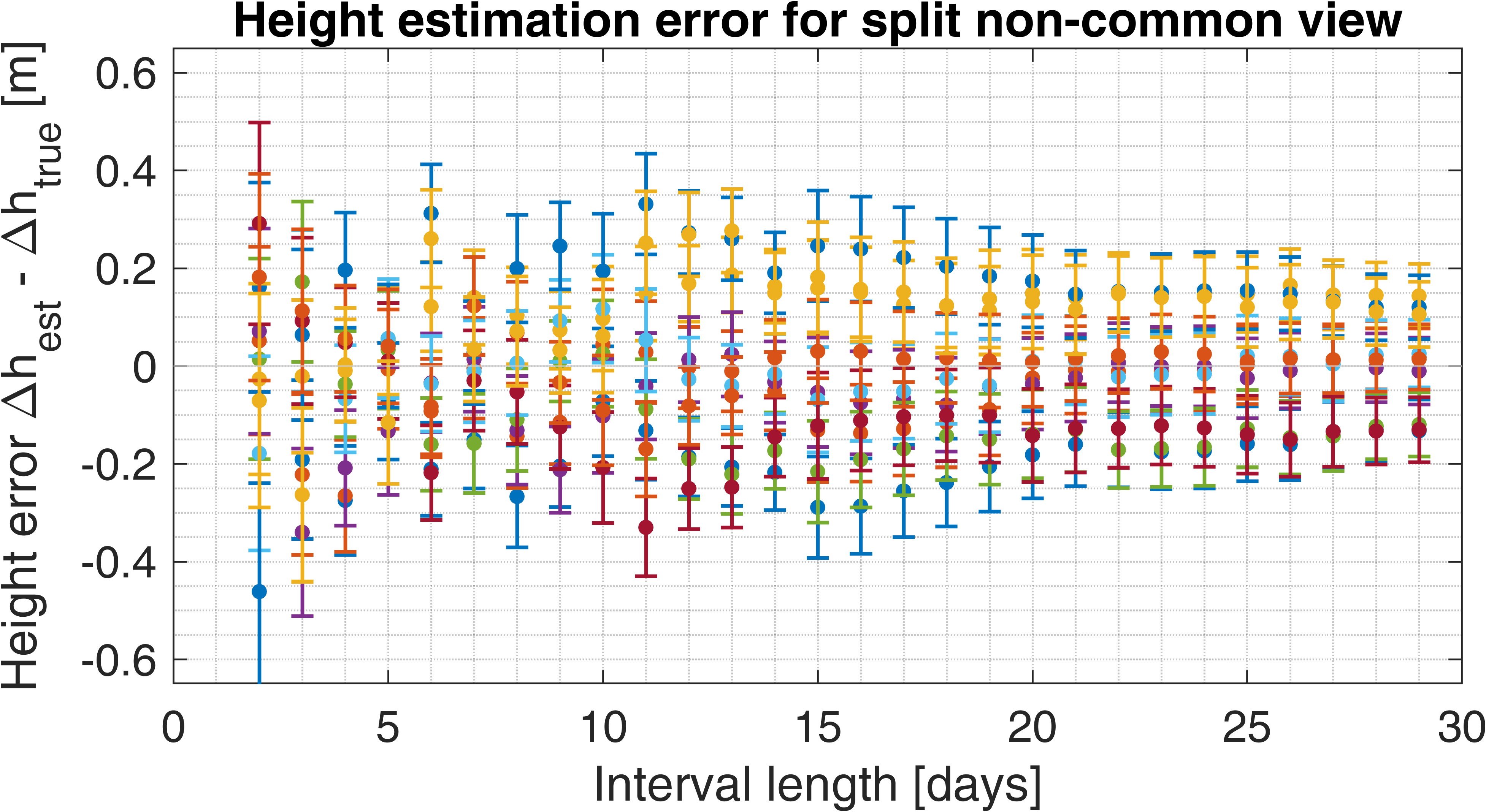}
			\caption*{(b) All station pairs}
		\end{minipage}
		\begin{minipage}{0.20\linewidth}
			\includegraphics[width=\linewidth]{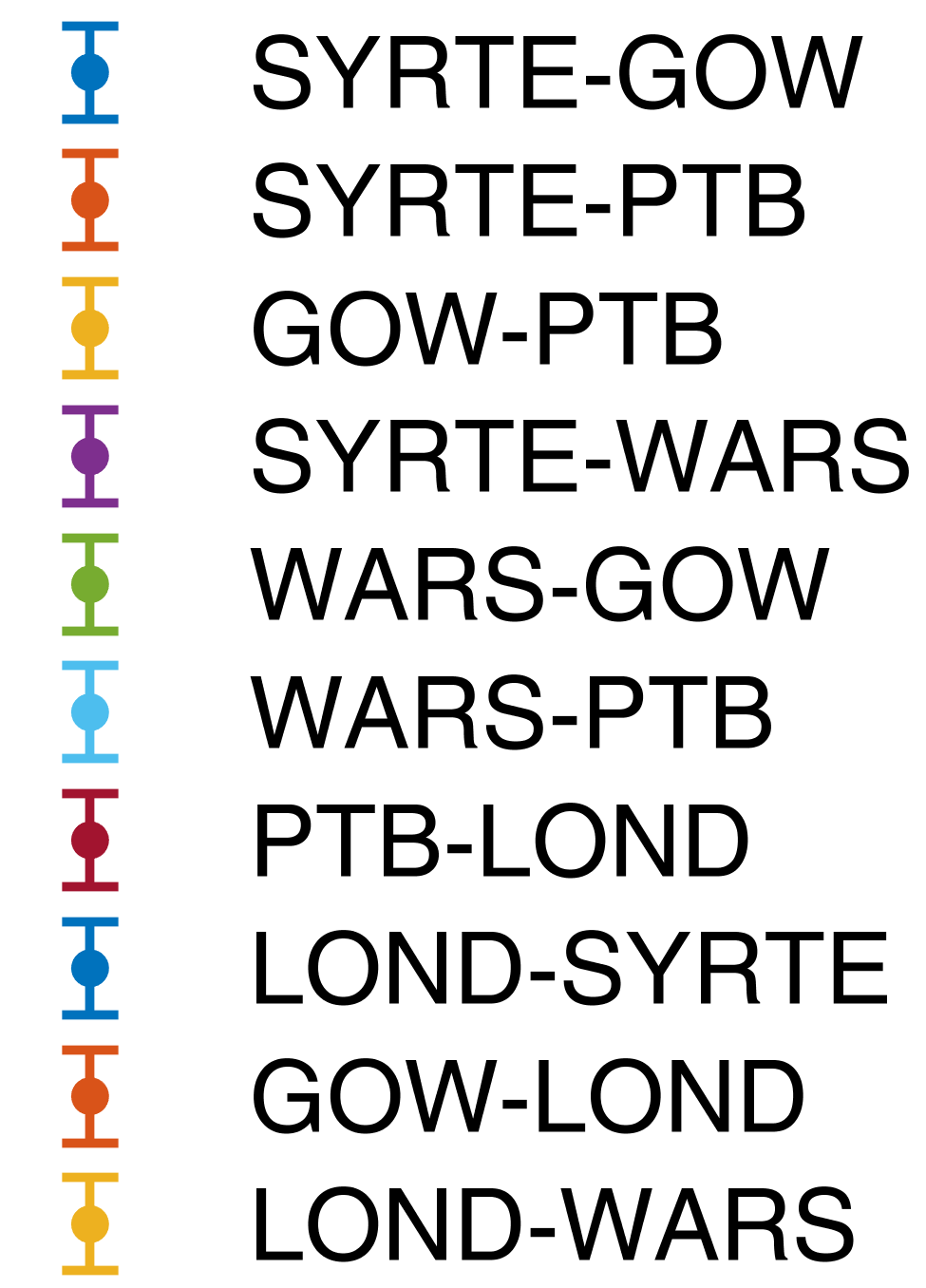}
            \vspace{10pt}
		\end{minipage}
		
	\end{minipage}
	
	\caption{Height estimation error for split non-common-view observations. In contrast to the previous figures, the legend is labeled according to the locations of the compared clocks rather than the ELT or MWL station names, as both link types are combined in this configuration. The clock locations therefore represent the actual physical height differences being estimated.}
	
	\label{fig:splitNCV_combined}
\end{figure}

The split non-common view results are shown in Figures~\ref{fig:splitNCV_combined}. In contrast to the previous approaches, both MWL and ELT observations are jointly used in the adjustment. The first solution is possible only after approximately two days, since at least two satellite passes per station are required to estimate local slopes within each interval.
For the three selected station pairs (Figure~\ref{fig:splitNCV_combined}(a)), the height estimation errors exhibit a comparatively smooth temporal evolution. Large systematic drifts, as observed in the standard non-common view MWL solutions, are no longer present. Instead, the estimates fluctuate around quasi-stationary bias levels that gradually decrease with increasing interval length.
This behavior reflects the two-stage estimation process. By subdividing the observation period into shorter intervals, long-term correlated errors affecting individual links are partly absorbed in the local slope estimates and do not propagate coherently into the final adjustment. As a result, the split non-common view approach softens systematic trends at the cost of increased statistical scatter.
The results for all station pairs (Figure~\ref{fig:splitNCV_combined}(b)) confirm this pattern. Despite the heterogeneous network geometry and different link characteristics, the solutions cluster within a relatively narrow error band. After about two weeks, most station pairs achieve height errors on the order of a few decimeters, with some approaching the decimeter level.
The joint use of MWL and ELT observations improves the temporal coverage and allows the inclusion of stations that would otherwise lack sufficient quasi-common view opportunities. However, the combined solutions remain less accurate than the best ELT-only results, indicating that the microwave observations introduce additional noise and bias contributions that are not fully averaged out in the combined adjustment.

Overall, the split non-common view method provides a robust compromise between observation availability and estimation accuracy. It enables the determination of height differences for a larger station network while maintaining stable solutions even in the absence of simultaneous visibility. Notably, already for interval lengths of two to three days, height errors in the low-decimeter range and in some cases even at the centimeter level, can be achieved. This early convergence is particularly relevant for practical applications, where continuous clock operation over long periods cannot always be guaranteed. This makes the approach suitable for realistic ACES observation scenarios with irregular pass distributions.

\section{Conclusion}
\label{sec:conclusion}
This study investigated the determination of physical height differences between distant locations using satellite-based clock comparisons in the context of the ACES mission. Using a realistic full-scale simulation, different observation geometries and time-transfer techniques were evaluated, including quasi-common view, non-common view, and split non-common view configurations based on microwave and optical links. 

The results demonstrate that optical time-transfer links provide significantly faster convergence and higher long-term stability than microwave links. In quasi-common view geometry, ELT-based solutions reach the decimeter level within a few days and can approach the centimeter level for longer observation periods. Microwave-based solutions, in contrast, remain affected by larger fluctuations and systematic biases due to temporally correlated error contributions.
Non-common view processing increases observation availability and enables height determination even without simultaneous satellite visibility. While the accuracy is typically slightly degraded compared to quasi-common view, the overall behavior of the individual links remains similar, indicating that link-specific noise characteristics continue to dominate the solutions. 
The split non-common view approach further improves the practical applicability of the method by subdividing the observation period into shorter intervals and combining the resulting local slope estimates. This strategy reduces long-term correlated errors and enables the inclusion of stations with irregular pass availability. Although the combined MWL and ELT solutions are less accurate than the best ELT-only results, they provide stable estimates for a significantly larger network. In particular, interval lengths of two to three days already yield height differences at the decimeter or even centimeter level for several station pairs.
From an operational perspective, these short intervals are especially relevant, since continuous long-term operation of optical clocks cannot always be assumed. The results therefore indicate that meaningful geopotential information can be obtained even from relatively short observation campaigns. 

Overall, the results show that satellite-based clock comparisons using optical links constitute a promising remote-sensing technique for relativistic geodesy. The methods investigated here provide a solid basis for the analysis of forthcoming ACES observations and demonstrate the potential for determining physical height differences on a continental scale without relying on ground-based fiber connections. Future work will focus on the application of these methods to real measurement data and on further improving robustness under operational conditions. In particular, future simulation studies should also account for the use of flywheel oscillators to maintain time continuity during clock interruptions, as demonstrated in the realization of optical timescales \cite{bib_gre_2016}. Furthermore, effects such as correlations between observations, the presence of outliers, and varying clock and link performance across different station pairs should be investigated, as well as observation limitations due to weather-dependent constraints such as cloud coverage,  which specifically influence the availability of ELT measurements. Accounting for these aspects is expected to yield more realistic performance estimates for operational scenarios.

\medskip
\noindent{\small
\textbf{Author Contributions:} Conceptualization, J.M. and K.E.L.; Methodology, K.E.L., J.M., P.V. and A.S.; Software, P.V. and K.E.L.; Validation, K.E.L. and J.M.; Formal Analysis, K.E.L.; Investigation, K.E.L. and P.V.; Resources, P.V. and A.S.; Data Curation, K.E.L and P.V.; Writing – Original Draft Preparation, K.E.L. and P.V.; Writing – Review \& Editing, K.E.L., P.V., J.M. and A.S.; Visualization, K.E.L. and P.V.; Supervision, J.M.; Project Administration, J.M.; Funding Acquisition, J.M. and A.S.

\medskip

\noindent\textbf{Funding:} This research was funded by the Deutsche Forschungsgemeinschaft (DFG, German Research Foundation), Project-ID 490990195 (FOR 5456: Clock Metrology: A Novel Approach to TIME in Geodesy).

\medskip

\noindent\textbf{Data Availability Statement:} The data are available from the corresponding author upon reasonable request.

\medskip

\noindent\textbf{Conflicts of Interest:} The authors declare no conflict of interest.
}

\isPreprints{}{%
\begin{adjustwidth}{-\extralength}{0cm}
    }

\reftitle{References}




\isAPAandChicago{}{%

}

\isChicagoStyle{%

}{}

\isAPAStyle{%

}{}

%


\PublishersNote{}
\isPreprints{}{
\end{adjustwidth}
} 
\end{document}